\documentclass[twocolumn,nofootinbib,showpacs,preprintnumbers,amsmath,amssymb,floatfix]{revtex4}

\usepackage{graphicx}
\usepackage{dcolumn}
\usepackage{bm}
\usepackage{amssymb}
\usepackage{amsmath}
\usepackage{hep}
\usepackage{color}
\usepackage{multirow}

\preprint{UB-ECM-PF-10/10}
\preprint{CFTP/10-004}

\newcommand{\hzero}{\ensuremath{\PHiggslightzero}} 
\newcommand{\Hzero}{\ensuremath{\PHiggsheavyzero}} 
\newcommand{\Azero}{\ensuremath{\PHiggspszero}} 
\newcommand{\Hpm}{\ensuremath{\PHiggspm}} 
\newcommand{\CP}{\ensuremath{\mathcal{C}\mathcal{P}}}
\newcommand{\lag}{\ensuremath{\mathcal{L}}}
\newcommand{\jump}{\vspace{0.3cm}}

\newcommand{\retildehat}{\ensuremath{\hat{\Sigma}}}
\newcommand{\retilde}{\ensuremath{\Re e\,\Sigma}}
\newcommand{\self}{\ensuremath{\Sigma}}
\newcommand{\sw}{\ensuremath{s_W}}
\newcommand{\cw}{\ensuremath{c_W}}
\newcommand{\swd}{\ensuremath{s_W^2}}

\newcommand{\phzero}{\ensuremath{\HepProcess{\APelectron\Pelectron \HepTo \hzero\PZ^0}}}
\newcommand{\pHzero}{\ensuremath{\HepProcess{\APelectron\Pelectron \HepTo \Hzero\PZ^0}}}

\begin{document}

\hyphenation{Higgs-boson quantum-effects two-Higgs-doublet-model
des-cri-be li-te-ra-tu-re par-ti-cu-lar a-na-ly-sis
in-ves-ti-ga-ting si-tua-ti-on com-for-ta-ble ty-pi-ca-lly
tri-go-no-me-tric}

\title{Quantum effects on Higgs-strahlung events at Linear Colliders  \\ within
the general 2HDM}

\author{David L\'opez-Val}
 \email{dlopez@ecm.ub.es}
\author{Joan Sol\`a}%
 \email{sola@ecm.ub.es}
\affiliation{%
High Energy Physics Group, Dept. ECM, and Institut de Ci{\`e}ncies del Cosmos\\
Univ. de Barcelona, Av. Diagonal 647, E-08028 Barcelona, Catalonia, Spain}%

\author{Nicol\'as Bernal}
\email{nicolas.bernal@cftp.ist.utl.pt} \affiliation{\textit{Centro de
F\'isica Te\'orica de Part\'iculas (CFTP), Instituto Superior
T\'ecnico, Avenida Rovisco Pais, 1049-001 Lisboa, Portugal}}

\date{\today}

\begin{abstract}
The associated production of neutral Higgs bosons with the $\PZ^0$
gauge boson ($\hzero\PZ^0$, $\Hzero\PZ^0$) is investigated in the
context of the future linear colliders, such as the ILC and CLIC,
within the general two-Higgs-doublet model (2HDM). We compute the
corresponding cross-section for the processes $\APelectron\Pelectron
\to \PZ^0\,\hzero\,/\PZ^0\,\Hzero$ at one-loop, including the full
set of corrections at $\mathcal{O}(\alpha^3_{ew})$ together with the
leading  $\mathcal{O}(\alpha^4_{ew})$ terms, in full consistency
with the available theoretical and phenomenological constraints. We
find that the wave-function corrections to the external Higgs fields
are the dominant source of the quantum effects, which turn out to be
large and negative (e.g. $\delta\,\sigma/\sigma \sim -20\%\,/
-60\%$) in all the $\sqrt{s}$ range, and located predominantly in
the region around $\tan\beta \sim 1$ and moderate values of the
parameter $\lambda_5$ (being $\lambda_5 < 0$). This behavior can be
ultimately traced back to the enhancement potential of the triple
Higgs boson self-couplings, a trademark feature of the 2HDM with no
counterpart in the Higgs sector of the Minimal Supersymmetric
Standard Model. Even under this substantial depletion of the
one-loop corrected signal (which is also highly distinctive with
respect to the SM expectation for $\APelectron\Pelectron \to
\PZ^0\,\PHiggs$), the predicted Higgs-strahlung rates comfortably
reach a few tens of femtobarn, which means barely $\sim \,10^3 -
10^4$ events per $500$ $\invfb$ of integrated luminosity. Due to
their great complementarity, we argue that the combined analysis of
the Higgs-strahlung events ($\hzero\PZ^0$, $\Hzero\PZ^0$) and the
previously computed one-loop Higgs-pair production processes
($\hzero\Azero$, $\Hzero\Azero$) could be instrumental to probe the
structure of the Higgs sector at future linac facilities.
\end{abstract}


\maketitle

\vskip 6mm

\section{Introduction}
\label{sec:intro}

After more than 40 years since the seminal ideas coined by a handful
of theoretical pioneers\,\cite{pioners}, our present understanding
of the Electroweak Symmetry Breaking (EWSB) phenomenon through the
Higgs (Englert-Brout and Guralnik-Hagen-Kibble) mechanism is still
rather incomplete and experimentally inconclusive. On the one hand,
there is no compelling alternative to consistently embed the EWSB
mechanism into the quantum field theoretical description of particle
physics offered by the -- in so many regards successful -- Standard
Model (SM) of the Strong and Electroweak interactions. On the other
hand, we do not have a single phenomenological hint of the existence
of elementary scalar fields, not to mention the fact that we do not
understand how to make compatible the EWSB mechanism and its
associated vacuum energy with fundamental problems of different
scope, for example in the domain of cosmology. Still, the
possibility to describe the inner theoretical structure of the SM
through the EWSB mechanism is so successful within the restricted
particle physics domain that it would not be wise, not even
advisable, to cease our pursue of the phenomenological implications
of the Higgs mechanism paradigm till the limits of the current
experimental possibilities.

It goes without saying that the quest for experimental evidences of
the Higgs boson is a most preeminent milestone of the upcoming
generation of collider facilities. Nonetheless, the future data
might well reveal that the purportedly found Higgs boson actually
belongs to a richer model structure, which might be grounded
somewhere beyond the minimal conception of the SM, namely of a
single, spinless, fundamental constituent of matter. If so, a few
more fundamental spinless constituents could appear. A particularly
well-motivated extension is the two-Higgs-doublet model structure
encompassed by the Minimal Supersymmetric Standard Model
(MSSM)\,\cite{MSSM}. Here the physical Higgs boson spectrum contains
a couple of \CP-even (\hzero, \Hzero), one \CP-odd (\Azero) and two
charged Higgs bosons ($\PHiggs^\pm$), and the corresponding Higgs
potential is highly constrained by the underlying Supersymmetry
(SUSY). One particular consequence of the latter is that SUSY
invariance restricts all Higgs boson self-interactions to be
gauge-like. While this represents an economy from the point of view
of the number of couplings in the potential, it is strongly
imbalanced by the exceeding number of parameters (mixings and
masses) populating the other sectors of the MSSM. In addition, the
pure gauge nature of the Higgs boson self-couplings make them highly
inconspicuous from the practical point of view, in the sense that
they are unable to trigger any outstanding phenomenological
signature. The core of the enhancement capabilities of the MSSM
Lagrangian resides, instead, in the multifarious pattern of Yukawa
couplings between the Higgs bosons and the quarks, as well as
between quarks, squarks and charginos/neutralinos. The rich
interplay of opportunities that they give rise to has been
extensively analyzed in the past within a plethora of processes, see
e.g.~\cite{Coarasa:1996qa,Jimenez:1995wf,GHS0298,CarenaGNW,BGGS02},
and also ~\cite{Carena:2002es,anatomy,Heinemeyer:2008tw,hunter} for
reviews on the subject.

The two-Higgs $SU_L(2)$-doublet structure of the Higgs sector in the
MSSM constitutes a genuine prediction of the SUSY dynamics.
Nonetheless, a more general architecture can appear in the form of a
non-SUSY framework through the so-called general (unconstrained)
two-Higgs-doublet Model (2HDM)\,\footnote{To be more precise, we
refer to this model as ``general'' in the sense that we allow all
possible operators leading to a renormalizable, gauge-invariant and
\CP-conserving Higgs potential. As usually done in
practice\,\cite{hunter}, we impose an additional (softly-broken)
$\mathcal{Z}_2$ discrete symmetry $\Phi_i \to (-1)^i\,\Phi_i$, where
$i = 1,2$ denote each of the $SU_L(2)$ Higgs doublets, as a
sufficient condition to banish the tree-level flavor changing
neutral currents.}. Although they share a common Higgs boson
spectrum, the potentially most distinctive
phe\-no\-me\-no\-lo\-gi\-cal features of both models are located in
very different sectors. The most relevant observation here is that,
in the absence of an underlying SUSY, the triple (3H) and quartic
(4H) Higgs-boson self-interactions are no longer restrained to be
purely gauge. This can have a tremendous impact e.g. in the physics
of the top quark in hadron colliders, as it was shown long ago in
\cite{Coarasa:1998xy}, and it can also trigger significant neutral
flavor-changing interactions\,\cite{Bejar:2000ub} that can perfectly
compete with the SUSY ones\,\cite{Guasch:1999jp,Bejar:2008ub}.

Much attention has been devoted to Higgs boson production and decay
in hadron colliders, extending from the ongoing Tevatron facility at
Fermilab to the brand new LHC collider recently operating at CERN\,
\cite{Carena:2002es,anatomy,Heinemeyer:2008tw,hunter,Coarasa:1998xy,Bejar:2000ub,Guasch:1999jp,Bejar:2008ub,Draper:2009fh,Arhrib:2009hc}.
However, precision Higgs boson physics will greatly benefit from the
interplay\,\cite{Weiglein:2004hn} of the forthcoming generation of
linear colliders (linac), such as the ILC and CLIC
projects\,\cite{linacs}. Here a variety of processes can provide new
clues to Higgs boson physics, e.g. the production of triple
Higgs-boson final states, both in the MSSM\, \cite{Djouadi:1999gv}
and in the general 2HDM \cite{giancarlo}; the double Higgs-strahlung
channels $hh\PZ^0$ \cite{arhrib}; and the inclusive Higgs-pair
production via gauge-boson fusion\,\cite{Djouadi:1999gv,neil}. In
the same vein, also the $\Pphoton\Pphoton$ mode of a linac has been
explored\,\cite{Krawczyk:2003hb}, in particular the loop-induced
production of a single neutral Higgs boson \cite{nicolas} and of a
Higgs boson pair \cite{twophoton}. In all the above mentioned cases,
promising signatures were singled out and illustrate that, if
effectively realized in Nature, such hints of a generic 2HDM
structure could hardly be missed in the superbly clean environment
of a linac. And, what is more, they could not be confused as having
a SUSY origin, because of the intrinsically different nature of the
Higgs self-interaction sector. Besides, outstanding fingerprints of
a generic 2HDM could also be stamped in the pattern of radiative
corrections to Higgs production processes; for instance, quantum
effects on the cross-sections for two-body Higgs boson final states:
\begin{eqnarray}
  \APelectron\Pelectron \to 2h\,\ \ \ \ \ (2h \equiv \hzero\,\Azero;\, \Hzero\,\Azero;\,
\PHiggsplus\PHiggsminus)\,, \label{2H}
\end{eqnarray}
\noindent have been attentively investigated in the MSSM
\cite{mssmloop1,mssmloop2,mssmloop3,mssmloop4}. As for the general
2HDM, the efforts were first concentrated on the production of
charged Higgs pairs \cite{2hdm_charged}. This program has been
recently brought to completion from a full-fledged study of the
quantum effects on the production cross-sections in the neutral
Higgs sector \cite{main}. Alongside these processes we find the more
traditional Higgs-strahlung events, in which a Higgs boson is
produced in association with the $\PZ^0$:
\begin{equation}
\APelectron\Pelectron \to h\,\PZ^0,\ \ \ h = \hzero, \Hzero\,.
\label{eq:hz}
\end{equation}
(Notice that the $\Azero \PZ^0$ final state is forbidden by
$\CP$-conservation.) Processes (\ref{eq:hz}) are complementary to
the (\ref{2H}) ones -- cf. \cite{sumrules}, and also~\cite{pairmssm,
Djouadi:1999gv} for a phenomenological analysis in the MSSM context.
Let us also recall that the last available limit on the SM Higgs
boson mass, placed by LEP searches, comes precisely from
investigating the ``Bjorken process''\,\cite{Bjorken76}, i.e.
$\APelectron\Pelectron \to {\rm H}\,\PZ^0$ followed by $\PZ^0\to
f\bar{f}$, with the result: $M_H\gtrsim 114.4$
GeV\,\cite{Amsler:2008zzb}. In this article, we aim at discussing
the corresponding one-loop corrections to the generalized Bjorken
processes \eqref{eq:hz} within the 2HDM. However, in contrast to the
original SM case, where the radiative corrections are small, the
quantum effects on the processes (\ref{eq:hz}) can be large and are
mainly driven by the 3H self-couplings. In fact, our main aim here
is to identify the regions of the parameter space where this is so,
and then quantify the impact of the potentially enhanced 3H
self-couplings on the final cross-sections. In this way, joining
this study with that of Ref.~\cite{main} on the pairwise Higgs boson
production channels (\ref{2H}) in the 2HDM, a rather complete
panorama of the genuine quantum effects associated to the basic
neutral Higgs boson production channels becomes available.

\section{Higgs-strahlung events at one-loop: theoretical setup}
\label{sec:theo}

A good deal of attention has been devoted in the literature to
multiple properties of the Higgs-strahlung process
$\APelectron\Pelectron \to \PHiggs \PZ^0$ in the SM, see
e.g.~\cite{zh_sm} and references therein. Not surprisingly it was
one of the gold-plated channels for Higgs boson search at LEP.
Concerning the MSSM, the Higgs-strahlung channels ~\eqref{eq:hz}
have also been discussed extensively -- cf.
Refs.~\cite{mssmloop3,zh_mssm} and part two of the
review~\cite{anatomy} -- and are currently under investigation also
in the MSSM with \CP-violating phases\,\cite{zh_cp} (the so-called
complex MSSM\,\cite{Frank:2006yh}). The upshot of these analyses
spotlights the following features: i) the dominant source of
corrections at one-loop originate from the Higgs boson propagators,
which can be reabsorbed into an \emph{effective} (loop-corrected)
mixing angle $\alpha_\text{eff}$ (equivalently, a more generic
mixing matrix in the {complex MSSM} case); ii) the corrections to
the $\PZ\PZ\,h$ vertex are in general small, although they can reach
the level of $10\,\%$ for very low (or high) values of $\tan\beta$,
precisely in the regions where the Higgs Yukawa couplings to heavy
quarks become enhanced; iii) the electromagnetic corrections to the
initial state with virtual photonic corrections and initial-state
radiation do not differ from the SM case, being in general large and
positive (except near the production threshold). In the present
article, our endeavor is to seek for the genuine phenomenological
imprints associated to the generic (unconstrained) 2HDM dynamics,
most particularly to the potentially enhanced 3H self-couplings.

Let us recall that the general 2HDM\,\cite{hunter} is obtained by
canonically extending the SM Higgs sector with a second $SU_L(2)$
doublet with weak hypercharge $Y=+1$, so that it contains $4$
complex scalar fields. The free parameters $\lambda_i$ in the
general, \CP-conserving, 2HDM potential can be finally expressed in
terms of the masses of the physical Higgs particles ($M_{h^0}$,
$M_{H^0}$, $M_{A^0}$, $M_{H^\pm}$), $\tan \beta$ (the ratio of the
two VEV's $\langle H_i^0\rangle$ giving masses to the up- and
down-like quarks), the mixing angle $\alpha$ between the two
$\CP$-even states and,  last but not least, the self-coupling
$\lambda_5$, which cannot be absorbed in the previous quantities.
Therefore we end up with a $7$-free parameter set, to wit:
$(M_{h^0}$, $M_{H^0}$, $M_{A^0}$, $M_{H^{\pm}}$, $\sin\alpha$,
$\tan\beta$, $\lambda_5)$. Furthermore, to ensure the absence of
tree-level flavor changing neutral currents (FCNC), two main 2HDM
scenarios arise: 1) type-I 2HDM, in which one Higgs doublet couples
to all quarks, whereas the other doublet does not couple to them at
all; 2) type-II 2HDM, where one doublet couples only to down-like
quarks and the other doublet to up-like quarks. The MSSM Higgs
sector is actually a type-II one, but of a very restricted sort
(enforced by SUSY invariance)\,\cite{hunter}. We refer the reader to
Ref. \cite{main} for a comprehensive account on the structure of the
model and for notational details. In particular, Table II of that
reference includes the full list of trilinear couplings within the
general 2HDM that are relevant for the present calculation. Further
constraints must be imposed to assess that the SM behavior is
sufficiently well reproduced up to the energies explored so far.
Most particularly, we take into account i) the approximate $SU(2)$
custodial symmetry, which can be reshuffled into the condition
$|\delta\rho_{2HDM}|\le 10^{-3}$\,\cite{custodial}; and $ii)$ the
agreement with the low-energy radiative $B$-meson decay data (which
demand $M_{H^{\pm}} \gtrsim 300$ GeV for $\tan \beta \ge 1$
\cite{bound_charged} in the case of type-II 2HDM).

{Additional requirements} ensue from the theoretical consistency of
the model, to wit i) perturbativity ~\cite{perturbativity}; ii)
unitarity~\cite{unitarity}; and iii) stability of the 2HDM
vacuum~\cite{vacuum}. Let us expand a bit more on the latter
conditions, which turn out to play a crucial role in our analysis.
As for the unitarity bounds, and following Ref.~\cite{unitarity},
the basic underlying strategy is to compute the S-matrix elements
$S_{ij}$ for the possible $2 \to 2$ processes involving Higgs and
Goldstone bosons in the 2HDM, and to subsequently restrain the
corresponding eigenvalues $U_{ik}\,S_{kl}U^{-1}_{lj} =
\alpha_i\,\delta_{ij}$ by the generic condition $|\alpha_i| < 1/2 \,
\forall i$. The latter requirement translates into a set of upper
bounds on the quartic scalar couplings $\lambda_i, \, i = 1 \dots 6$
-- and hence on combinations of Higgs masses, trigonometric
couplings and, most significantly, the parameter $\lambda_5$. As far
as vacuum stability restrictions are concerned \cite{vacuum}, they
may be written as follows:
\begin{eqnarray}
&& \lambda_1 + \lambda_3 > 0\, ; \quad \lambda_2 + \lambda_3 > 0 \nonumber \\
&& 2\sqrt{(\lambda_1 + \lambda_3)(\lambda_2 + \lambda_3)} +
2\lambda_3 + \lambda_4
\nonumber \\
&&+ \mbox{min}[0, \lambda_5 - \lambda_4, \lambda_6 - \lambda_4] > 0
\label{eq:vacuum_conditions}.
\end{eqnarray}
\noindent More refined versions of these equations may be obtained
from the Renormalization Group (RG) running of the quartic couplings
$\lambda_i(\mu^2)$ at high energies. Nonetheless, for our current
purposes we need not assume any particular UV completion of the
2HDM. This would be an unnecessary additional assumption
at this stage of the phenomenological analysis of our processes.
Therefore, we are not tied to any specific UV-cutoff
(which could be, in principle, as
low as $\Lambda \sim 1-10\,\TeV$). In this sense, we
may just apply Eq.~\eqref{eq:vacuum_conditions} with all $\lambda_i$
taken at the EW scale. This is after all the scale at which
we fix all our input parameters and perform the
phenomenological analysis (including the renormalization)
of the processes under study.
By the same token, the (cutoff dependent) triviality bounds are also
circumvented. The latter kind of restrictions only impose tight
upper limits on the Higgs boson masses when a very large UV-cutoff
(e.g. the Planck scale or the GUT scale) is considered. Our choices
of Higgs boson mass spectra (cf. Table~\ref{tab:masses}) lie, in any
case, in a sufficiently low mass range so as to conform with the
typical mass requirements allowed by triviality~\cite{triviality} --
even for large cutoff scales. More details on the constraints setup
are provided in Ref.~\cite{main}.

\begin{figure*}[thb]
\begin{center}
\begin{tabular}{c}
\includegraphics[scale=1.0]{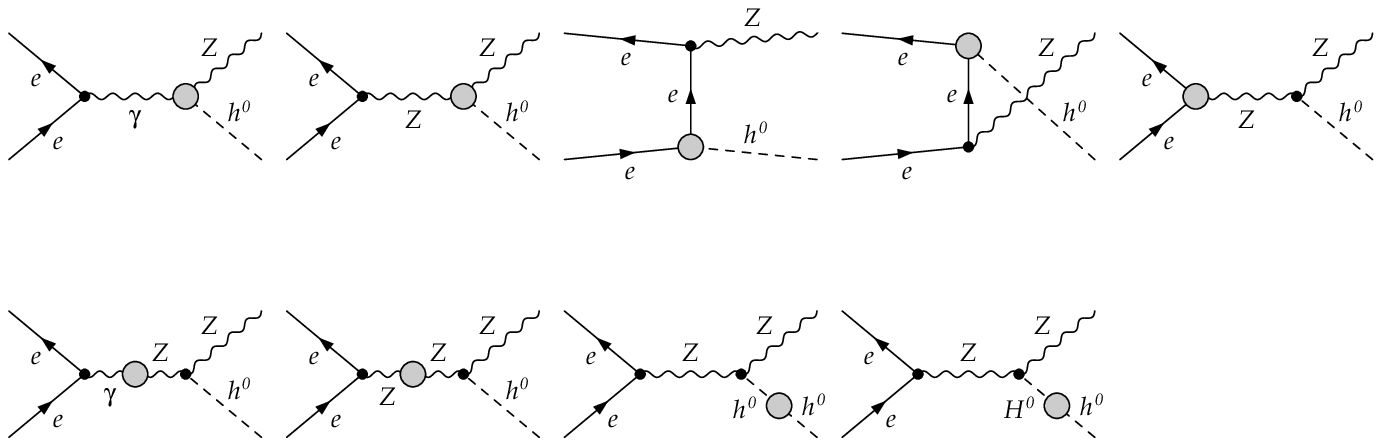}  \\ \quad \\ \\ \quad \\
\includegraphics[scale=1.0]{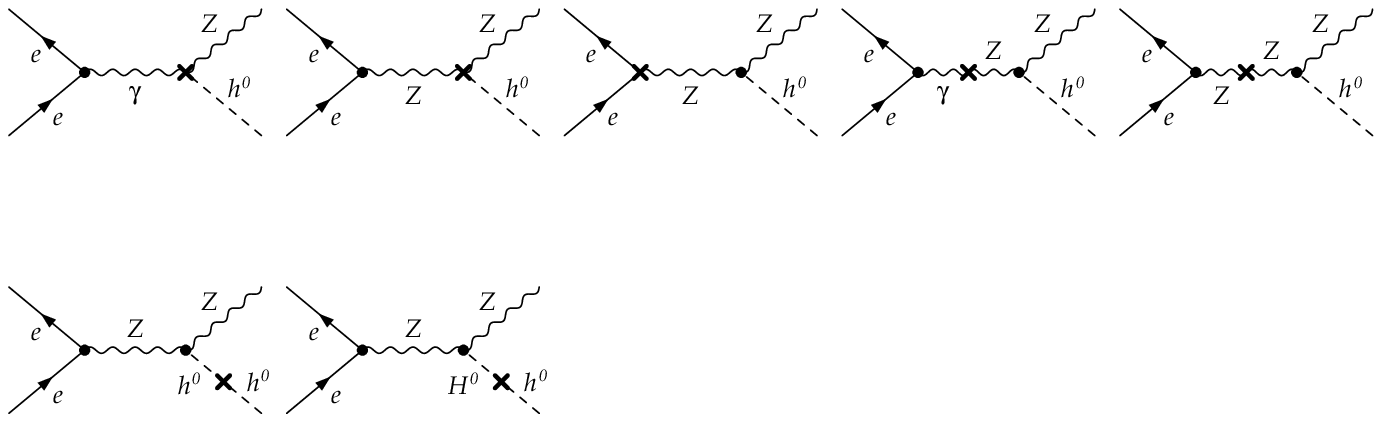}  \\ \quad \\
\end{tabular}
\caption{Set of Feynman diagrams contributing to
$\APelectron\Pelectron \to \PHiggslightzero\PZ^0$ at one-loop level
within the 2HDM. The shaded blobs stand for all possible loops with
virtual particles. An equivalent collection of diagrams accounts for
the complementary process $\pHzero$. The last couple of rows
displays the complete set of one-loop counterterm diagrams.}
\label{fig:loop}
\end{center}
\end{figure*}

\jump
The dynamics of the Higgs-strahlung processes under consideration is driven
at leading order by the tree-level interaction Lagrangians:
\begin{eqnarray}\label{eq:lag0}
\lag_{\PZ^0\PZ^0\,\hzero} &=&
\frac{e\,\sin(\beta-\alpha)\,M_Z}{\sw\,\cw}\,g^{\mu\nu}\,\PZ^0_\mu\,\PZ^0_\nu\,\hzero\,,\nonumber\\
\lag_{\PZ^0\PZ^0\,\Hzero} &=&
\frac{e\,\cos(\beta-\alpha)\,M_Z}{\sw\,\cw}\,g^{\mu\nu}\,\PZ^0_\mu\,\PZ^0_\nu\,\Hzero\,,
\end{eqnarray}
where $\sw\equiv\sin\theta_W$ and $\cw\equiv\cos\theta_W$ for the
electroweak mixing angle $\theta_W$. Since
$\lag_{\PHiggspszero\,\PHiggslightzero\,\PZ^0}\propto
\cos(\beta-\alpha)$ and
$\lag_{\PHiggspszero\,\PHiggsheavyzero\,\PZ^0}\propto\sin(\beta-\alpha)$,
it is clear that the processes (\ref{eq:hz}) are complementary to
the Higgs pair production ones (\ref{2H}). Furthermore, as all these
couplings (\ref{eq:lag0}) are generated by the gauged kinetic terms
of the Higgs doublets (cf. Eq.~(30) of Ref.~\cite{main}), they are
fully determined by the gauge symmetry and hence show no intrinsic
difference in the 2HDM as compared to the MSSM. At the end of the
day, this is the reason why a tree-level analysis of these events is
most likely insufficient to disclose their true nature. Similarly,
in the limit $\alpha = \beta - \pi/2$ the $\hzero\PZ^0\PZ^0$
coupling coincides with the analogue coupling in the SM,
$\PHiggs\PZ^0\PZ^0$. It is, therefore, the pattern of radiative
corrections the characteristic signature associated to each one of
the possible models; most particularly, it should help to
disentangle SUSY versus non-SUSY extended Higgs physics scenarios.

The leading-order $\mathcal{O}(\alpha_{ew})$ scattering amplitude follows
from the $s$-channel $\PZ^0$-boson exchange and renders
\begin{eqnarray}
&& {\cal M}^{(0)}(\phzero) = -\frac{e^2\,M_Z\,\sin(\beta-\alpha)}{\sw\,\cw\,(s-M_Z^2)}\, \times \nonumber \\
&& \times
\bar{v}(p_1,\eta_1)\,\slashed{\epsilon}\,(k_2,\sigma_2)\,(g_L\,P_L +
g_R\,P_R)\,u(p_2,\eta_2)\,.\nonumber \\
\label{eq:amp0}
\end{eqnarray}
Here $p_{1,2}$ and $\eta_{1,2}$ refer to the 4-momenta and
helicities of the electron and positron, and
${\epsilon}\,(k_2,\sigma_2)$ is the polarization 4-vector of the $Z$
gauge boson with 4-momentum $k_2$ and helicity $\sigma_2$. We have
introduced also the left and right-handed weak couplings of the
$\PZ^0$ boson to the electron, ${g_L} = (-1/2 + \swd)/\cw\sw$,
${g_R} = \sw/\cw$, and the left and right-handed projectors
$P_{L,R}=(1/2)(1\mp\gamma_5)$.
Let us notice that we do not include the finite $\PZ^0$-width
corrections, since they are completely negligible for the
center-of-mass energies that we consider here. Finally, the total
cross section $\sigma(\phzero)$ at the tree-level is obtained after
squaring the matrix element~\eqref{eq:amp0}, performing an averaged
sum over the polarizations of the colliding $\APelectron\Pelectron$
beams and the outflowing $\PZ^0$ boson, and integrating over the
scattering angle.

The calculation of $\sigma(\phzero)$ at one loop is certainly much
more cumbersome. To start with, it is UV-divergent and it thus
requires of a careful renormalization procedure in order to render
finite results. We adopt here the conventional on-shell scheme in
the Feynman gauge ~\cite{renorm_sm}. In the MSSM, these
cross-section calculations can be mostly carried out in an
automatized fashion through standard algebraic packages which allow
a rapid and efficient analysis of the electroweak precision
observables\,\cite{Heinemeyer:2004gx}. Several public codes are
available, see
e.g.\,\cite{feynarts,feynHiggs,AutomatRen,Baro:2008bg}. However, in
our case the calculation is non-supersymmetric and we must deal with
the renormalization of the Higgs sector in the class of generic 2HDM
models. A detailed description of the renormalization procedure for
the 2HDM Higgs sector in the on-shell scheme has been presented in
\cite{main} and we refer the reader to this reference for all the
necessary details. We will make constant use of the framework
described in this reference and sometimes we will refer to
particular formulae of it.

\begin{figure}[thb]
\begin{center}
\begin{tabular}{c}
\includegraphics[scale=0.5]{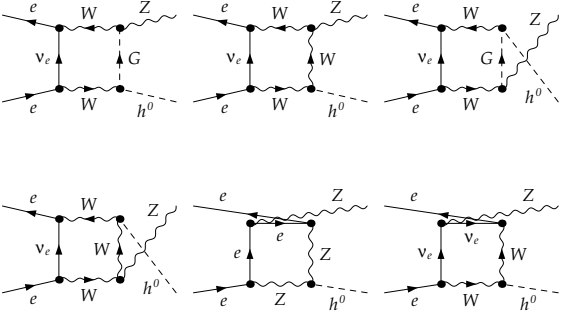} \hspace{1cm}
\end{tabular}
\caption{Set of Feynman diagrams contributing to
$\APelectron\Pelectron \to \PHiggslightzero\PZ^0$ at one-loop level
within the 2HDM. These diagrams describe the box-type quantum
corrections. An equivalent collection of diagrams accounts for the
complementary process $\pHzero$.} \label{fig:box}
\end{center}
\end{figure}
\begin{figure}[thb]
\begin{center}
\begin{tabular}{c}
\includegraphics[scale=0.55]{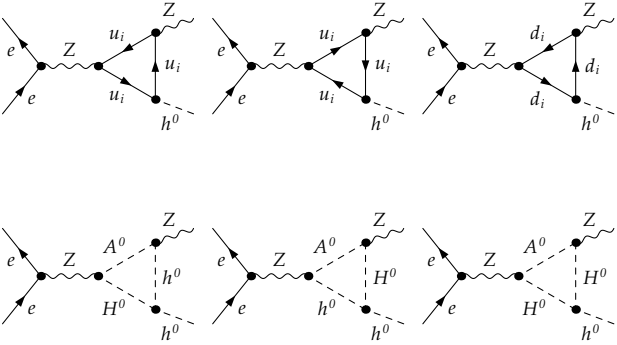} \hspace{1cm}
\end{tabular}
\caption{Sample of Feynman diagrams describing the one-loop
corrections to the $\hzero\PZ^0\PZ^0$ interaction for the process
$\phzero$ within the 2HDM, including quark-mediated (top line) and
Higgs boson-mediated quantum corrections (bottom line).}
\label{fig:vertex}
\end{center}
\end{figure}

\begin{figure}[thb]
\begin{center}
\begin{tabular}{c}
\includegraphics[scale=0.55]{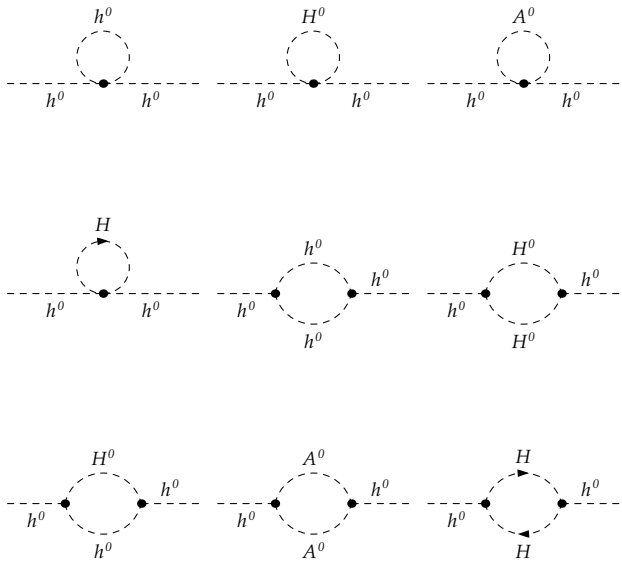} \hspace{1cm}
\end{tabular}
\caption{Subset of Feynman diagrams driving the dominant
contribution to the $\hzero$ boson self-energy. Notice that the
two-point functions are sensitive to two triple Higgs-boson
self-couplings. The tadpole diagrams, however, which depend on
quartic couplings, do not contribute to the WF renormalization. An
equivalent collection of diagrams contribute to the corresponding
$\Hzero$ self-energy.} \label{fig:self}
\end{center}
\end{figure}

With this renormalization setup in mind, the various contributions
to the scattering amplitude of $\APelectron\Pelectron \to
\PHiggslightzero\PZ^0$ at one-loop (Figs.
\ref{fig:loop}-\ref{fig:box}) can be classified in a meaningful way.
First of all, let us note that all of the one-loop diagrams are of
course of order $\mathcal{O}(\alpha^2_{ew})$. However, some of them
include, in addition, enhancement factors sourced by the trilinear
Higgs boson couplings $\lambda_{3H}$, see Table II of \cite{main}.
In such cases, we shall include these factors when assessing the
order of magnitude of the diagram. We are now ready for sorting out
the one-loop contributions in different categories:

\begin{itemize}
\item{Self-energy corrections to the $\PZ^0$
and $\Pphoton-\PZ^0$ mixing propagators, all of them of order
$\mathcal{O}(\alpha^2_{ew})$ with no enhancement factors at this
order.}
\item{Vertex corrections to the $\APelectron\Pelectron \PZ^0$ interaction,
which are also of order $\mathcal{O}(\alpha^2_{ew})$. It should be
noted that we do not include the virtual photonic
$\mathcal{O}(\alpha_{em}\, \alpha_{ew})$ effects nor the real
bremsstrahlung emission off the $\APelectron/\Pelectron$ legs. These
pure QED corrections and the weak ones factorize into two subsets
which are separately UV finite and gauge-invariant. Moreover, these
photonic contributions are fully insensitive, at the order under
consideration, to the relevant 3H self-couplings on which we focus.
For this kind of processes involving electrically neutral Higgs
bosons in the final state, the one-loop QED effects are confined to
the initial $\APelectron\Pelectron$ vertex. In practice, the net
outcome of the accompanying initial state radiation (ISR) is to
lower the effective center-of-mass energy available for the
annihilation process. All in all, these effects are unessential at
this stage to test the presence of the new dynamical features
triggered by the 2HDM in the Higgs-strahlung events under analysis.}
\item{Vertex corrections to the $\APelectron\Pelectron\PHiggs$ interaction.
Since we have explicitly set $m_e = 0$ throughout our calculation,
the $\APelectron\Pelectron\PHiggs$ tree-level Yukawa coupling is
absent and the corresponding one-loop diagrams automatically render
a UV finite contribution which is, in any case, very small.}
\item{The loop-induced $\Pphoton\PZ^0\,\hzero$ interaction. This one
is order $\mathcal{O}(e\,\alpha_{ew}\,\lambda_{3H})$ and therefore
includes an enhancement factor $\lambda_{3H}$ . Due to the
$\Pphoton-\PZ^0$ mixing at one-loop, the following counterterm is
needed so as to render a UV-finite vertex:
\begin{eqnarray}
\delta\,\lag_{\hzero\PZ^0\Pphoton} &=&
\frac{e\,\sin(\beta-\alpha)\,M_Z\,}{2\sw\cw}\,\delta\,Z_{\Pphoton\PZ^0}\,
g^{\mu\nu}\,\PZ^0_\mu\,A_\nu\,\hzero\, . \nonumber \\
\label{eq:zphotonlight}
\end{eqnarray}
\noindent and similarly for the counterterm associated to the
effective $\lag_{\Hzero\PZ^0\Pphoton}$} interaction:
\begin{eqnarray}
\delta\,\lag_{\Hzero\PZ^0\Pphoton} &=& \frac{e
\,\cos(\beta-\alpha)\,M_Z} {2\sw\cw}\,\delta\,Z_{\Pphoton\PZ^0}\,
g^{\mu\nu}\,\PZ^0_\mu\,A_\nu\,\Hzero\, , \nonumber \\
\label{eq:zphotonheavy}
\end{eqnarray}

\item{
The vertex correction for $h\PZ^0\PZ^0$ encompasses different
contributions of order
$\mathcal{O}(\alpha^2_{ew},\,\alpha_{ew}\,e\,\lambda_{3H})$ (see the
sample diagrams in Fig.~\ref{fig:vertex}). And, most important,
there are, in addition, corrections of
$\,\mathcal{O}(\alpha_{ew}\,\lambda^2_{3H})$ (in fact, the dominant
ones) which are characterized by the highly conspicuous enhancement
factor squared $\lambda^2_{3H}$. The latter originates from the
associated vertex counterterm, most particularly from the Higgs
field renormalization constant
$Z_{\hzero}^{1/2}=1+(1/2)\delta\,Z_{\hzero}$. As we shall discuss in
more detail below, $\delta\,Z_{\hzero}$ is sensitive to the
scalar-scalar self-energy and hence it involves products of two
triple Higgs self-couplings (see Fig.~\ref{fig:self}). Of course, an
equivalent discussion holds for the complementary channel,
$\PHiggsheavyzero\,\PZ^0$.

The full form of the associated one-loop vertex counterterms ensues
from the usual splitting of bare fields and parameters into the
renormalized ones and associated counterterms, to wit:
\begin{equation}\label{split}
\lag\left(g_0=g+\delta g,\phi_0=Z_i^{1/2}\phi\right) \to
\lag(g,\phi) + \delta\,\lag\,.
\end{equation}
For the $\hzero\PZ^0\PZ^0$ and $\Hzero\PZ^0\PZ^0$ case one gets
respectively:
\begin{widetext}
\begin{eqnarray}\label{eq:counter11}
 \hspace{-.7cm}\delta\,\lag_{\hzero\PZ^0\PZ^0}&=&\frac{e\,\sin\,(\beta-\alpha)\,M_Z}{\sw\,\cw}\,
 \Bigg[
    \frac{\sw^2 - \cw^2}{\cw^2}\,\frac{\delta \sw}{\sw}+\sin\beta\cos\beta\,\cot{(\beta-\alpha)}\,
    \frac{\delta\,\tan\beta}{\tan\beta}   \nonumber \\
& & + \frac{\delta\,e}{e}+ \frac{\delta\,M_Z^2}{2\,M_Z^2}+ \frac{1}{2}\,\delta\,Z_{\PHiggslightzero} + \frac{1}{2}\,\delta\,Z_{\PZ^0} +
   \frac{1}{2}\,\cot{(\beta-\alpha)}\,\delta Z_{\PHiggsheavyzero\PHiggslightzero}\Bigg]\,g^{\mu\nu}\,\PZ^0_\mu \PZ^0_\nu \PHiggslightzero
\,,
\end{eqnarray}
\begin{eqnarray}
 \hspace{-.7cm}\delta\,\lag_{\Hzero\PZ^0\PZ^0}&=&\frac{e\,\cos\,(\beta-\alpha)\,M_Z}{\sw\,\cw}\,
 \Bigg[
    \frac{\sw^2 - \cw^2}{\cw^2}\,\frac{\delta \sw}{\sw}- \sin\beta\cos\beta\,\tan{(\beta-\alpha)}\,\frac{\delta\,\tan\beta}{\tan\beta}
     \nonumber \\
&& +\frac{\delta\,e}{e} + \frac{\delta\,M_Z^2}{2\,M_Z^2}+ \frac{1}{2}\,\delta\,Z_{\PHiggsheavyzero} + \frac{1}{2}\,\delta\,Z_{\PZ^0}
  +
   \frac{1}{2}\,\tan{(\beta-\alpha)}\,\delta Z_{\PHiggsheavyzero\PHiggslightzero}\Bigg]\,g^{\mu\nu}\,\PZ^0_\mu \PZ^0_\nu \PHiggsheavyzero
\,,
    \label{eq:counter12}
\end{eqnarray}
\end{widetext}
}
where we can spot in the structure of these formulae the presence of
the above mentioned {Higgs field} renormalization counterterms
$\delta\,Z_{\hzero}$ and $\delta\,Z_{\Hzero}$. The term
$\delta\,Z_{\hzero\Hzero}$ accounts for the WF mixing
$\hzero\leftrightarrow\Hzero$ emerging from the $\PZ^0 \PZ^0
\PHiggslightzero$ and $\PZ^0 \PZ^0\PHiggsheavyzero$ vertices.
\item{The finite {wave-function (WF)} correction to the external Higgs fields
$\hzero$ or $\Hzero$ (Fig.~\ref{fig:self}). These are related to the
fact that we have chosen the residue of the $\Azero$-propagator at
its pole to be one, and therefore there is no more freedom to make
the same choice for the other Higgs bosons. This entails a finite WF
renormalization correction, see Ref.\cite{main} for details.
These finite renormalization effects are of utmost importance in
this case, as they trigger leading contributions of order
$\mathcal{O}(\alpha_{ew}\,\lambda^2_{3H})$. They actually drive the
very bulk of the quantum effects (see below for a more detailed
discussion).}
\item{Finally, the box-type $\mathcal{O}(\alpha^2_{ew})$
diagrams (Figure~\ref{fig:box}), whose contribution is
non-negligible, in particular at large center-of-mass energies --
since they are not suppressed as $1/{s}$.}
\end{itemize}

The dominance of the WF corrections is a very characteristic feature
of the processes under analysis. It originates from a non-trivial
cancelation between the Higgs boson field counterterms, which appear
in two different pieces of the overall one-loop amplitude, though
with opposite signs. Let us further elaborate on this important
point. Without loss of generality, we concentrate on the
$\hzero\PZ^0$ channel. The complete set of contributions to the
scattering amplitude up to the one-loop level may be split in the
following manner:
\begin{eqnarray}
{\cal M}^{(0+1)}_{\phzero} &=& {\cal M}_{\phzero}^{(0)} + {\cal M}_{\phzero}^{(1)} +  \nonumber \\
&& + \delta\,{\cal M}_{\phzero}^{(1)} + {\cal M}_{\phzero}^{\mbox{WF}} \label{eq:amplitude1}
\end{eqnarray}
\noindent wherein the finite WF corrections to the external Higgs field are introduced as
\begin{eqnarray}
{\cal M}_{\phzero}^{\mbox{WF}} &=& (\sqrt{\hat{Z}_{\hzero}}-1)\,{\cal M}^{(0)}_{\phzero}
\nonumber \\
&&= -\frac{1}{2}\,\underbrace{\Re
e\,\retildehat'_{\hzero}(M^2_{\hzero})}_{\mathcal{O}(\lambda_{3H}^2)}\,{\cal
M}^{(0)}_{\phzero} + \mathcal{O}(\alpha^3_{ew}). \nonumber  \\
\label{eq:wf2}
\end{eqnarray}
At this point we are making explicit use of the on-shell
renormalization conditions defined in \cite{main}, in particular
$\Re e\,\retildehat_{\hzero\Hzero}(M^2_{\hzero}) = 0$. Moreover, we
only retain those contributions that are leading-order in the triple
Higgs self-couplings. Let us notice that the counterterm piece
$\delta\,{\cal M}^{(1)}$ is sensitive to
$\mathcal{O}(\lambda^2_{3H})$ effects through the WF renormalization
of the Higgs fields. More specifically, from the explicit expression
of the $\hzero$-field counterterm as a linear combination of the two
Higgs doublet counterterms (see Section IV of Ref.~\cite{main} for
details)
\begin{eqnarray}
\delta\,Z_{\hzero}=\sin^2\alpha\, \delta Z_{\Phi_1} +
\cos^2\alpha\,\delta
Z_{\Phi_2}=-\retilde'_{\Azero}(M^2_{\Azero}) + \dots \,,\phantom{yzk}
\label{eq:wf3}
\end{eqnarray}
\noindent we may single out the following
$\mathcal{O}(\lambda^2_{3H})$ contribution from Eq.~\eqref{eq:counter11},
\begin{eqnarray}
\delta\,\lag_{\hzero\PZ^0\PZ^0}&\to&
\frac{e\,\sin(\beta-\alpha)}{2\,\sw\,\cw}\,
\delta\,Z_{\hzero}+\, \dots \nonumber \\
&&= -\frac{e\,\sin(\beta-\alpha)}{2\,\sw\,\cw}\,
\underbrace{\retilde'_{\Azero}(M^2_{\Azero})}_{\mathcal{O}(\lambda_{3H}^2)}+\, \dots\, . \nonumber \\
\label{eq:demo1}
\end{eqnarray}
But, as warned, the finite WF renormalization factor of the $\hzero$
field in (\ref{eq:wf2}) is sensitive to
$\mathcal{O}(\lambda^2_{3H})$ terms too:
\begin{eqnarray}
\Re e\,\retildehat'_{\hzero}\,(M^2_{\hzero}) &=& \retilde'_{\hzero}\,(M^2_{\hzero})
+ \delta\,Z_{\hzero} \nonumber \\
&& = \underbrace{\retilde'_{\hzero}\,(M^2_{\hzero})
- \retilde'_{\Azero}\,(M^2_{\Azero})}_{\mathcal{O}(\lambda_{3H}^2)} + \dots\,, \nonumber \\
\label{eq:demo2}
\end{eqnarray}
\noindent where the dots stand for those terms with dependencies
other than $\mathcal{O}(\lambda^2_{3H})$. Notice, therefore, that
part of the $\mathcal{O}(\lambda^2_{3H})$ dependence cancels out
between the counterterm diagram associated to the $\hzero\PZ^0\PZ^0$
vertex -- cf. Eq.~\eqref{eq:demo1} -- and the finite $\hzero$
WF-factor -- cf. Eqs.~ \eqref{eq:wf2} and \eqref{eq:demo2}.
Specifically, it is the piece $ \retilde'_{\Azero}\,(M^2_{\Azero})$
that exactly cancels between the two terms. The remaining
$\mathcal{O}(\lambda^2_{3H})$ contributions come from
$\retilde'_{\hzero}(M^2_{\hzero})$ and are generated by the Feynman
diagrams displayed in Fig.~\ref{fig:self}. When sufficiently
enhanced, these pieces account for the bulk of the one-loop quantum
corrections. A typical diagram in the class of 2-point functions
provides the following contribution\,\footnote{Notice that we employ
the convention that $i\Sigma$ equals the Higgs boson self-energy
diagram, such that $\Sigma$ does not contain the global imaginary
part emerging from the loop integral. This is why we have multiplied
the loop integral in (\ref{eq:demo3}) by $-i$ before taking the real
part.}:

\begin{eqnarray}
&& {\cal M}_{\phzero}^{(1)}\simeq -\frac{1}{2}\,
\retilde'_{\hzero}(M^2_{\hzero})\,{\cal M}_{\phzero}^{(0)}\nonumber\\
&&\simeq -\frac{1}{2}\,{\cal M}_{\phzero}^{(0)}\,|\lambda_{3H}|^2\,
\times \nonumber \\
&&\times \Re e\frac{d^2}{dp^2}\,(-i)
\int\,\frac{d^D\,q}{(2\pi)^D}\,\frac{\mu^{4-D}}{(q^2 - M^2)\,
[(q+p)^2 - M^2]} \nonumber \\
&\sim&
-\frac{1}{2}\,\frac{|\lambda_{3H}|^2}{16\pi^2}\,{\cal M}_{\phzero}^{(0)}
\,\Re e B'_0(M^2_{\hzero},M^2,M^2)\, , \nonumber \\
\label{eq:demo3}
\end{eqnarray}
wherein the scalar two-point function is defined as in
Ref.~\cite{feynarts}:
\begin{eqnarray}
&& \int \frac{d^D\,q}{(2\pi)^D}\, \frac{\mu^{4-D}
}{(q^2-m_1^2)\,[(p+q)^2 - m_2^2]} \equiv\,
\nonumber \\
&& \quad \ \ \equiv \frac{i}{16\,\pi^2}\,B_0\,(p^2, m_1^2, m_2^2)\,,
\label{eq:2point}
\end{eqnarray}
$\mu$ denoting the 't Hooft mass unit. In the expression
(\ref{eq:demo3}), $M$ denotes the typical mass scale(s) which appear
in these one-loop 2-point functions. Owing to the derivative with
respect to the external momentum, the 1-point functions do not
contribute to the WF renormalization, and hence the quartic Higgs
boson self-couplings are not involved in this calculation. This
means that the first line of diagrams in Fig.~\ref{fig:self} does
not actually contribute.
We emphasize that the overall sign for this expression depends on
the sign of such 2-point function. Notice that $\Re\,e
B_0'(p^2,M^2,M^2)> 0$ for $p^2 < 4M^2$, which is the case we wish to
focus on for the $\hzero\PZ^0$ channel, and also for $\Hzero\PZ^0$
(as long as the resonant decay $\Hzero \to \hzero\hzero$ is
forbidden by kinematics). Therefore, in most of the scenarios of
interest, such finite WF corrections carry an overall minus sign.
Being proportional to $\lambda_{3H}^2$, they become the leading
quantum effects in the region where the trilinear couplings are
enhanced, and as a result they generate a characteristic pattern of
quantum effects in which a systematic suppression of the tree-level
cross-section is predicted.

The presence of the large negative corrections induced by the Higgs
boson self-energies brings forward a characteristic signature for
the production cross-sections of the Higgs-strahlung processes
(\ref{eq:hz}). This feature is in marked contradistinction to the
situation with the Higgs boson pair production mechanisms(\ref{2H}),
where the corresponding corrections are just opposite in sign, i.e.
large and positive, see \cite{main}. We shall further comment on
these interesting and correlated features in the next section. From
the general structure of Eq.\,(\ref{eq:demo3}), one may anticipate
the typical (maximum) size of the quantum effects on the
Higgs-strahlung processes as follows:
\begin{eqnarray}
\delta_r &=& \frac{\sigma^{(0+1)}-\sigma^{(0)}}{\sigma^{(0)}} =
\frac{\braket{2\,{\cal M}^{(0)}\,{\cal M}^{(1)}}}{\braket{|{\cal M}^{(0)}|^2}} \simeq
\nonumber \\
&& -\frac{|\lambda_{3H}|^2}{16\pi^2\,M^2}
\,f(M_{\hzero}^2,M^2,M^2)
\label{eq:estim},
\end{eqnarray}
$f$ being a dimensionless form factor (basically accounting for the
behavior of the $B'_0$ function). The notation $\braket{\dots}$
stands for the various operations of averaging and integration of
the squared matrix elements. Taking into account that unitarity
limits let the trilinear couplings reach values as large as
$\frac{\lambda_{3H}}{M_W} \simeq \frac{|\lambda_5|}{e} \simeq 30$,
and assuming $M \sim 200\,\GeV$ and $f \sim \mathcal{O}(1)$, the
above estimate typically predicts a strong depletion of the
tree-level signal by $\delta_r \simeq - 90\%$. This prediction falls
in the right ballpark of the exact numerical results that will be
reported in the next section, which point to a maximum depletion of
$\delta_r\sim -60\%$.

Before closing this section, let us recall that the counterterm
amplitude $\delta\,{\cal M}^{(1)}_{\phzero}$ derives from a number
of renormalization conditions that determine the renormalized
coupling constants and fields in a given renormalization framework.
As we have said, in our calculation the renormalization is performed
in the conventional on-shell scheme in the Feynman gauge,
appropriately extended to include the 2HDM Higgs sector. For the
latter, we need in particular a renormalization condition for the
parameter $\tan\beta=v_2/v_1$. We adopt the
following\,\cite{Dabelstein:1994hb}:
\begin{eqnarray}
 \frac{\delta\,v_1}{v_1} = \frac{\delta\,v_2}{v_2}
\label{eq:vacuumv1v2}.
\end{eqnarray}
This condition insures that the ratio ${v_2}/{v_1}$ is always
expressed in terms of the \emph{true} vacua after the
renormalization of the Higgs potential. The corresponding
counterterm resulting from $\tan\beta \rightarrow \tan\beta +
\delta\tan\beta$ can then be computed explicitly:
\begin{equation}\label{deltatanbetaexpl}
\frac{\delta\tan\beta}{\tan\beta}=\frac{1}{M_Z\,\sin 2\beta} \Re
e\self_{\PHiggspszero\,\PZ^0}(M^2_{\PHiggspszero})\,.
\end{equation}
This counterterm is involved in
Eqs.\,(\ref{eq:counter11})-(\ref{eq:counter12}), and it also
determines the WF mixing term $\delta\,Z_{\hzero\Hzero}$ that
appears in these equations, as follows: $\delta\,Z_{\hzero\Hzero}
=\sin{2\alpha}(\delta\tan\beta/\tan\beta)$. We refer the reader once
more to the exhaustive presentation of Ref.~\cite{main} for the
renormalization details.
%
\section{Numerical analysis}
\label{sec:numerical}
In this section, we present the numerical analysis of the one-loop
computation of the Higgs-strahlung processes \eqref{eq:hz} within
the 2HDM. We shall be concerned basically with the following two
quantities: i) the predicted cross-section at the Born-level
$\sigma^{(0)}$ and at one loop $\sigma^{(0+1)}$; and ii) the
relative size of the one-loop quantum corrections:
\begin{equation}
\delta_r = \frac{\sigma^{(0+1)}-\sigma^{(0)}}{\sigma^{(0)}}
\label{eq:deltar}.
\end{equation}
\begin{figure*}[t!]
\begin{center}
\begin{tabular}{ccc}
\includegraphics[scale=0.55,angle=0]{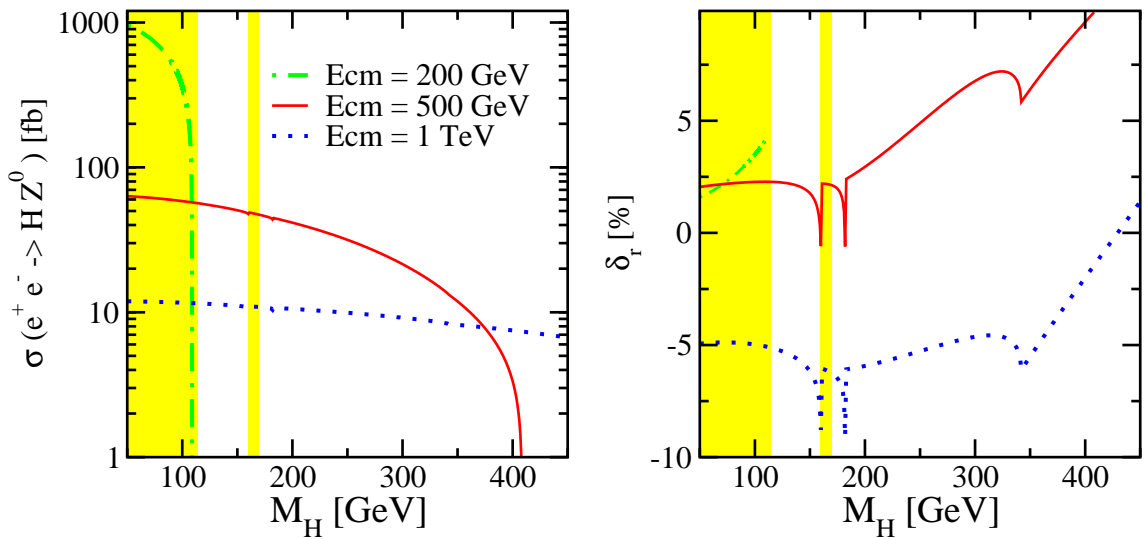} & \hspace{0.3cm} &
\end{tabular}
\caption{Total cross section $\sigma(\APelectron\Pelectron
\to \PHiggs\,\PZ^0)$ (in fb) at one-loop (left panel) and relative
one-loop correction $\delta_r$ (right panel) for $\sqrt{s}=500$ GeV
and $1$ TeV, as a function of $M_{\PHiggs}$ and within the SM. In
the leftmost side band we have computed also the curve corresponding
to LEP 200, which falls abruptly from rather high values down till
reaching the end of phase space. The thin vertical band corresponds
to the narrow exclusion region $160\lesssim M_H\lesssim 170$ GeV
determined by the Tevatron collaborations\,\cite{Krumnack:2009nz}.}
\label{fig:plotSM}
\end{center}
\end{figure*}
We have carried out our analysis with the help of the standard algebraic
and numerical tools \emph{FeynArts}, \emph{FormCalc} and \emph{LoopTools}
\cite{feynarts}.

The different Higgs boson mass sets that shall be used hereafter are
quoted in Table~\ref{tab:masses}. Let us highlight that, due to
their mass splittings (and also to the fact that $M_{\hzero}\geq
140$ GeV), Sets C and D can only be realized in a general (non-SUSY)
2HDM framework. Furthermore, in view of the value of the charged
Higgs mass, Sets A-C are only suitable for type-I 2HDM, whereas Set
D is valid for either type-I and type-II 2HDM's.

Apart from reflecting a variety of possible situations in the 2HDM
parameter space, some of these sets can be mimicked by the Higgs
boson mass spectrum in supersymmetric theories. For instance, Sets A
and B can be ascribed to characteristic benchmark scenarios of Higgs
boson mass spectra within the MSSM; in particular, Set B lies in the
class of the so-called maximal mixing scenarios~\cite{carena}, for
which $M_{\hzero}$ takes the highest possible values within the
MSSM. The numerical mass values for the MSSM-like sets have been
obtained with the aid of the program \emph{FeynHiggs} by taking the
full set of EW corrections at one-loop\,\cite{feynHiggs}.

We remark that the more massive the Higgs bosons are, the stronger
the constraints that unitarity imposes over $|\lambda_5|$. The
maximum (negative) values are roughly attained for $\lambda_5 \simeq
-9$ (Set A); $\lambda_5 \simeq -10$ (Sets B and C); and $\lambda_5
\simeq -8$ (Set D).
\begin{table}[h]
\begin{center}
\begin{tabular}{|c||c|c|c|c|}
\hline
 & $M_{\PHiggslightzero}\,[\GeV]$ & $M_{\PHiggsheavyzero}\,[\GeV]$ & $M_{\PHiggspszero}\,[\GeV]$ & $M_{\PHiggs^\pm}\,[\GeV]$  \\ \hline\hline
Set A   & $115$ & $220$ & $220$ & $235$  \\
Set B   & $130$ & $160$ & $150$ & $170$  \\
Set C   & $140$ & $150$ & $200$ & $200$  \\
Set D   & $150$ & $200$ & $260$ & $300$  \\
\hline
\end{tabular}
\caption{Choices of Higgs masses (in GeV) that are
used throughout the calculation. Due to the values of the
$M_{\PHiggs^{\pm}}$ mass, Sets A-C would only be suitable for type-I
2HDM, while Set D could account for both type-I and type-II models.
We also notice that Sets A,B have been devised in order to mimic the
characteristic mass splittings of the MSSM Higgs sector.}
\label{tab:masses}
\end{center}
\end{table}

Before coming to grips with the analysis of the Higgs-strahlung
events in the general 2HDM, it is interesting to briefly reconsider
the analogue process in the simpler framework of the SM. In
Fig.~\ref{fig:plotSM} we display the total (one-loop corrected)
cross section (left panel), together with the relative radiative
correction $\delta_r$ (right panel), as a function of the SM Higgs
boson mass $M_{\PHiggs}$. For completeness, and for illustration
purposes, we have also included the scenario corresponding to the
last Higgs boson mass segment ruled out by LEP at a center-of-mass
energy $\sqrt{s}=200$ GeV (see the leftmost vertical band in that
figure). The corresponding production rates for the ILC at
$\sqrt{s}=500$ GeV and $\sqrt{s}=1$ TeV are smaller than in the LEP
case due to the suppression of the s-channel amplitude by the
$Z$-boson propagator at higher energies, see Eq.\,(\ref{eq:amp0}),
and the larger mass of the produced Higgs boson. Still, the
cross-sections for producing SM Higgs bosons of a few hundred GeV at
the startup ILC energy ($\sqrt{s}=500$ GeV) lie at the
$40-60\,\femtobarn$ level, which leads roughly to $\sim 25,000$
Higgs events for an expected integrated luminosity of $500$
fb$^{-1}$.
In turn, the one-loop radiative corrections may be either
positive or negative, depending on the center-of-mass energy, and lie
generally at the level of a few percent, as can be seen in the right
panel of Fig.~\ref{fig:plotSM}. In this panel, we present the
evolution of the correction parameter  $\delta_r$ defined in
Eq.\,(\ref{eq:deltar}), as a function of $M_{\PHiggs}$. The three
peaks (also barely seen in the left panel) are correlated to the
production thresholds of $\PW^\PW^-$, $\PZ^0\PZ^0$ and $\Ptop\APtop$
pairs.

\begin{figure*}[t!]
\begin{center}
\begin{tabular}{ccc} \hspace{-0.3cm}
\includegraphics[scale=0.37,angle=-90]{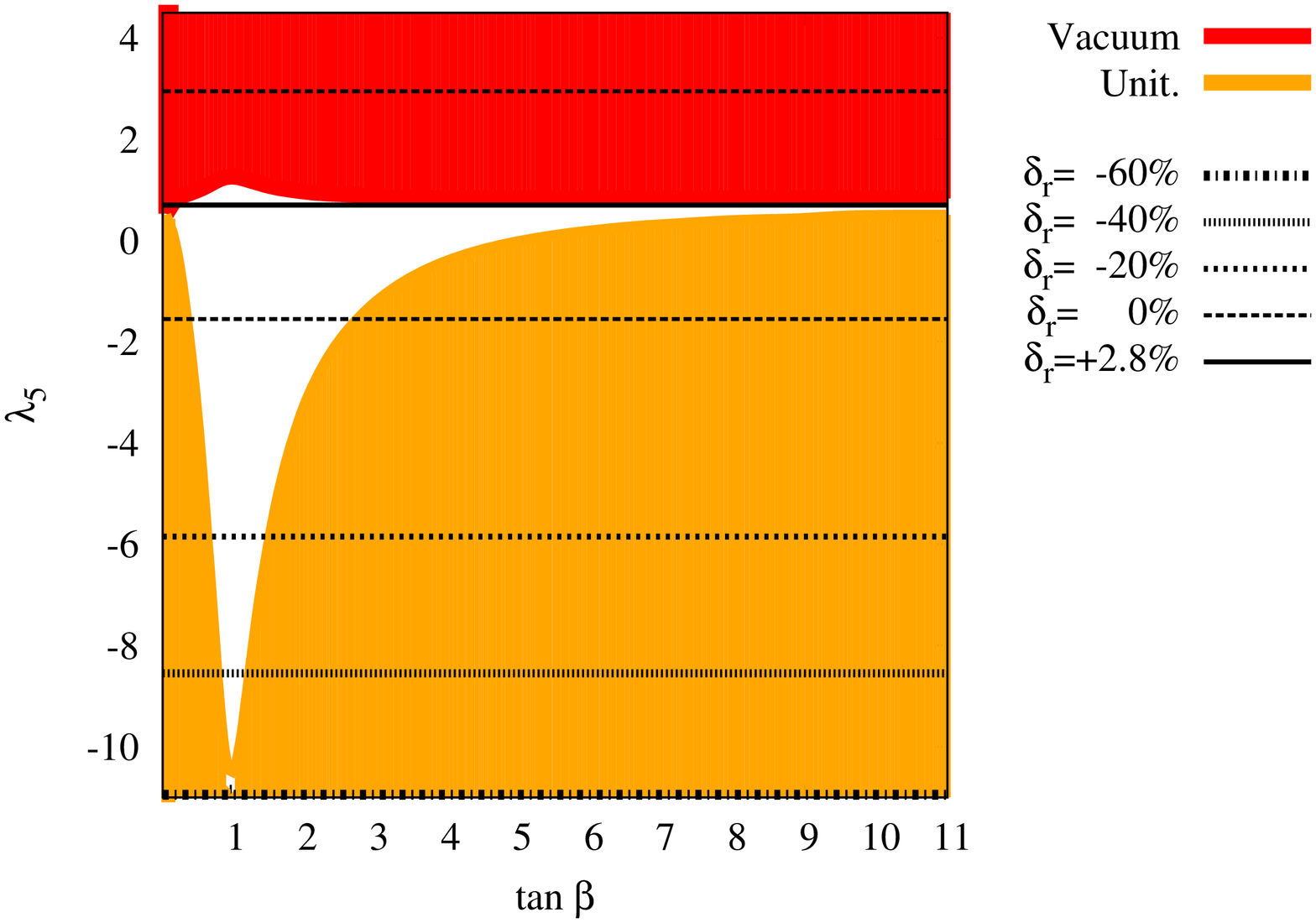} & \hspace{0.5cm} &
\includegraphics[scale=0.37,angle=-90]{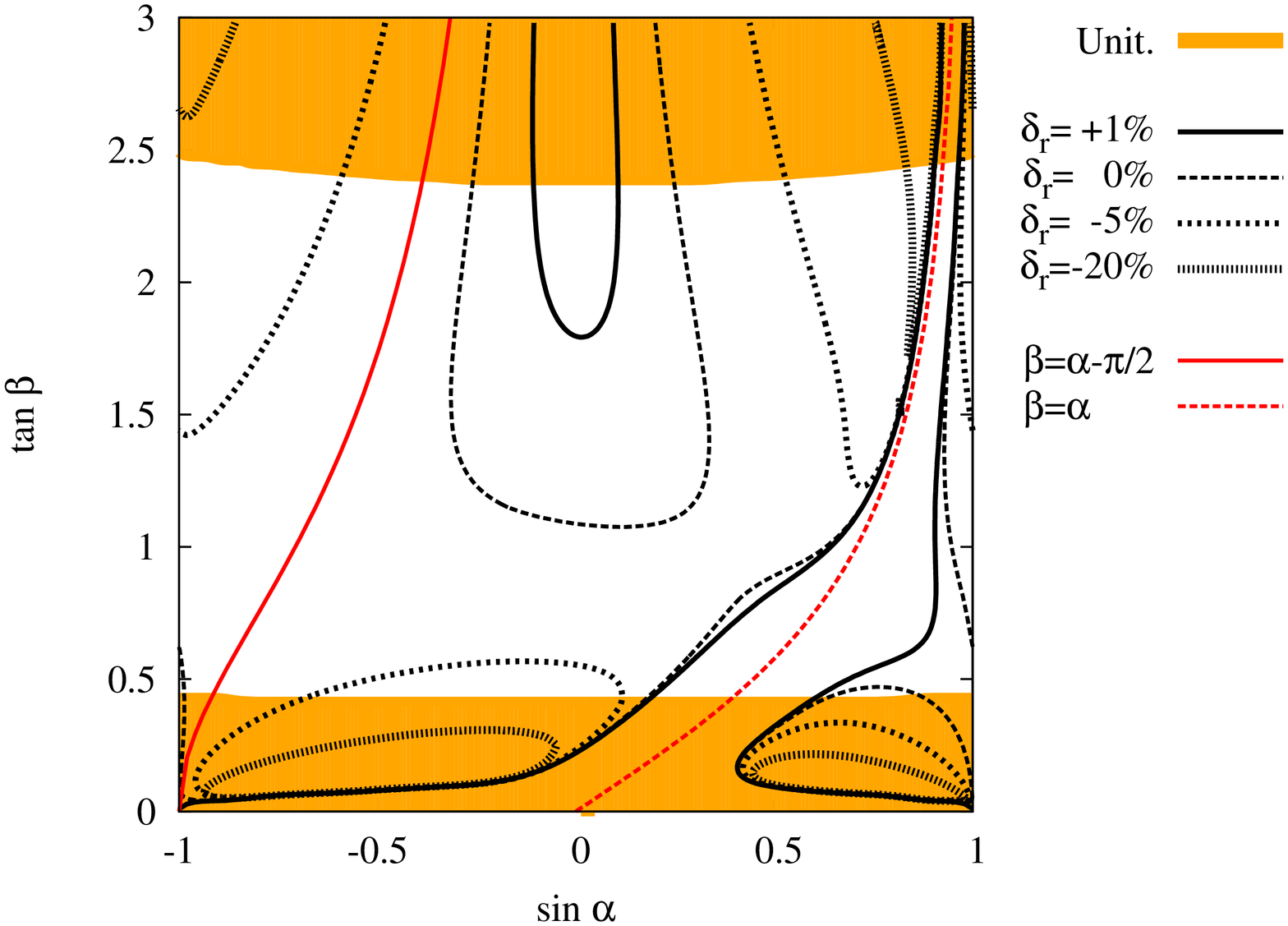}
\vspace{0.2cm}
\end{tabular}
\caption{Contour lines for the ratio $\delta_r$ in
the [$\tan\beta$, $\lambda_5$] plane assuming
$\alpha=\beta-\pi/2$ (left panel); similarly, in the
[$\sin\alpha$, $\tan\beta$] plane for $\lambda_5=-2$ (right
panel). The dark gray shaded areas stand for the regions excluded
by the vacuum stability bounds, whereas the light gray shaded
areas signal the domains excluded by the perturbative unitarity
bounds. The results are obtained for $\sqrt{s}=500$ GeV and the
Higgs boson masses as in Set B of Table~\ref{tab:masses}.}
\label{fig:iso}
\end{center}
\end{figure*}

\begin{figure*}[t]
\begin{center}
\begin{tabular}{ccc}
\includegraphics[scale=0.37,angle=-90]{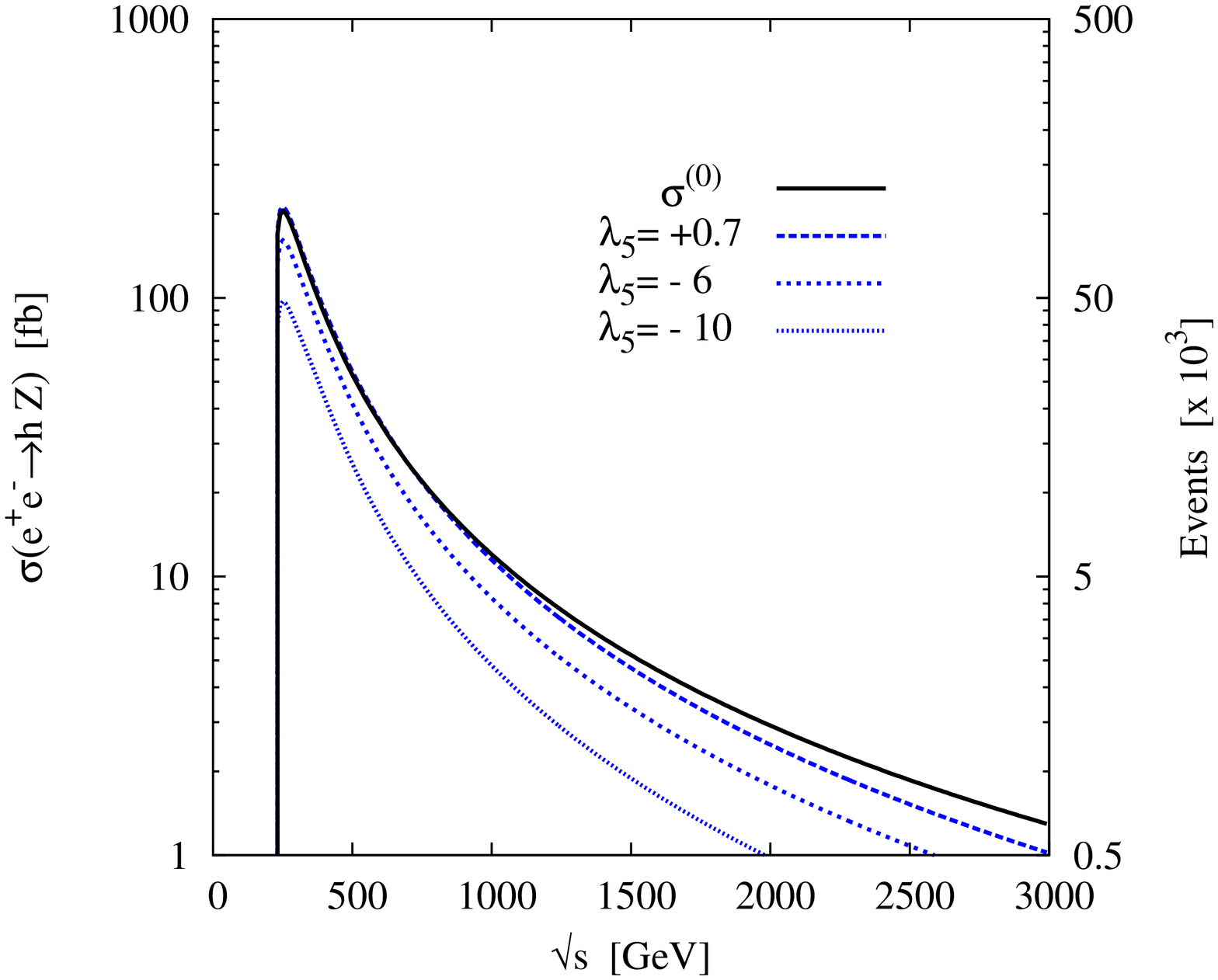} & \hspace{0.3cm} &
\includegraphics[scale=0.37,angle=-90]{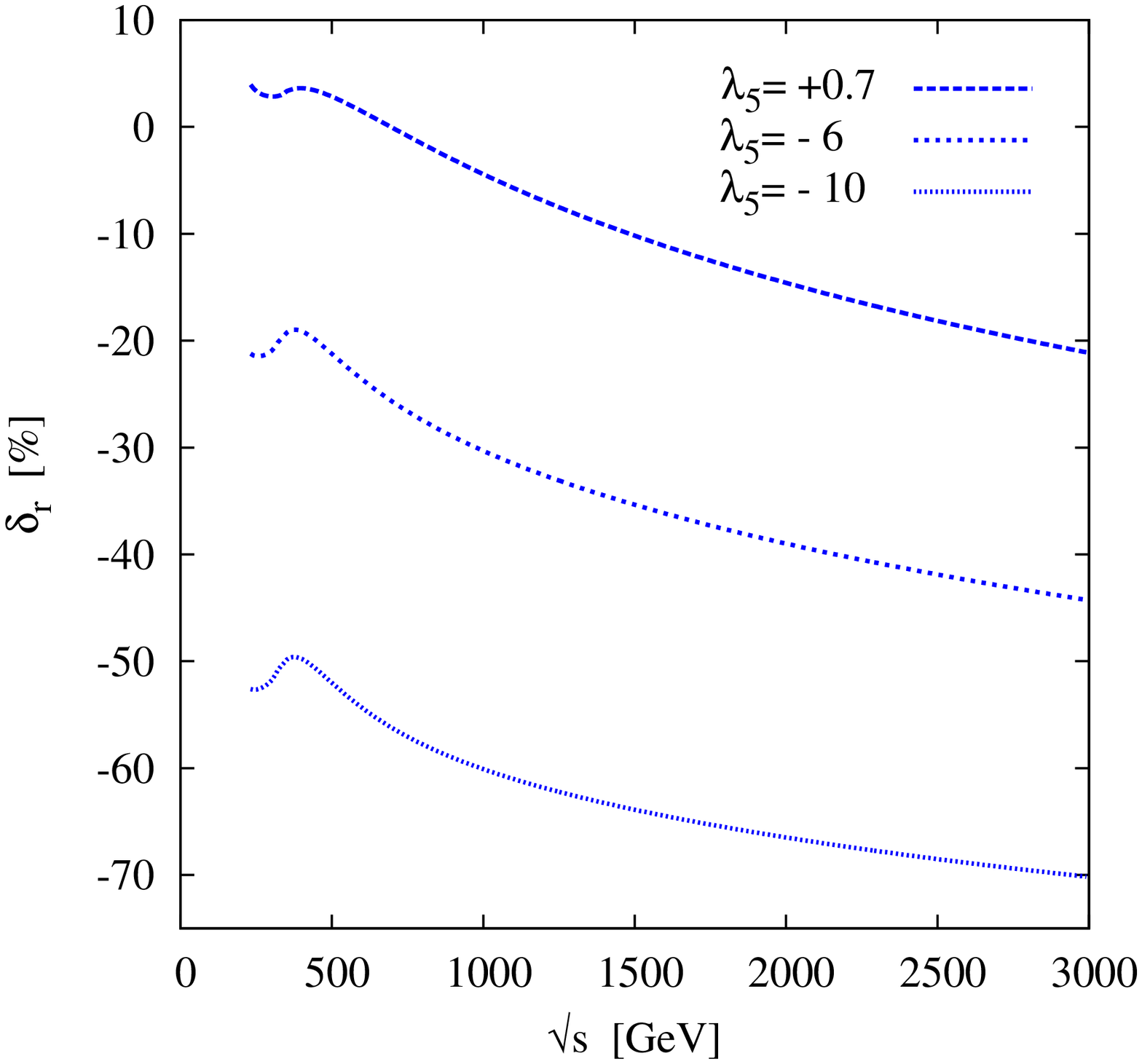} \\
&  & \\
\end{tabular}
\caption{Total cross section
$\sigma(\APelectron\Pelectron \to \PHiggslightzero\PZ^0)$ (in fb,
left panel) and relative one-loop correction $\delta_r$ (in $\%$,
right panel) as a function of $\sqrt{s}$ using Set B of Higgs boson
masses (cf. Table~\ref{tab:masses}); for $\tan\beta =1$, $\alpha =
\beta-\pi/2$ and three different values of $\lambda_5$. Shown are
also (cf. the right vertical axis of the left panel) the number of
events per $500$ $\invfb$ of integrated luminosity.}
\label{fig:plot3}
\end{center}
\end{figure*}

\begin{figure*}[t]
\begin{center}
\begin{tabular}{ccc}
\includegraphics[scale=0.37,angle=-90]{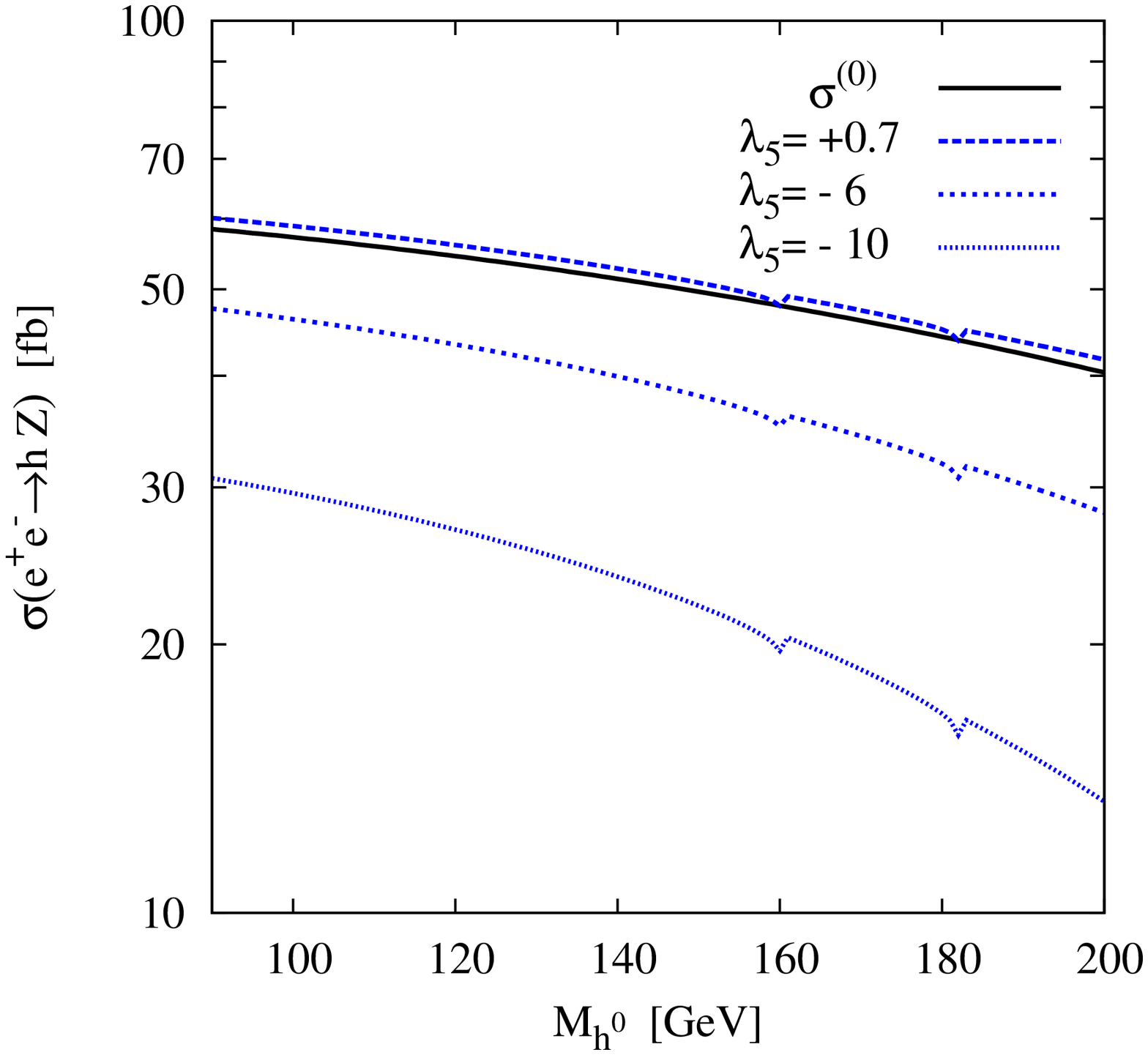} & \hspace{0.3cm} &
\includegraphics[scale=0.37,angle=-90]{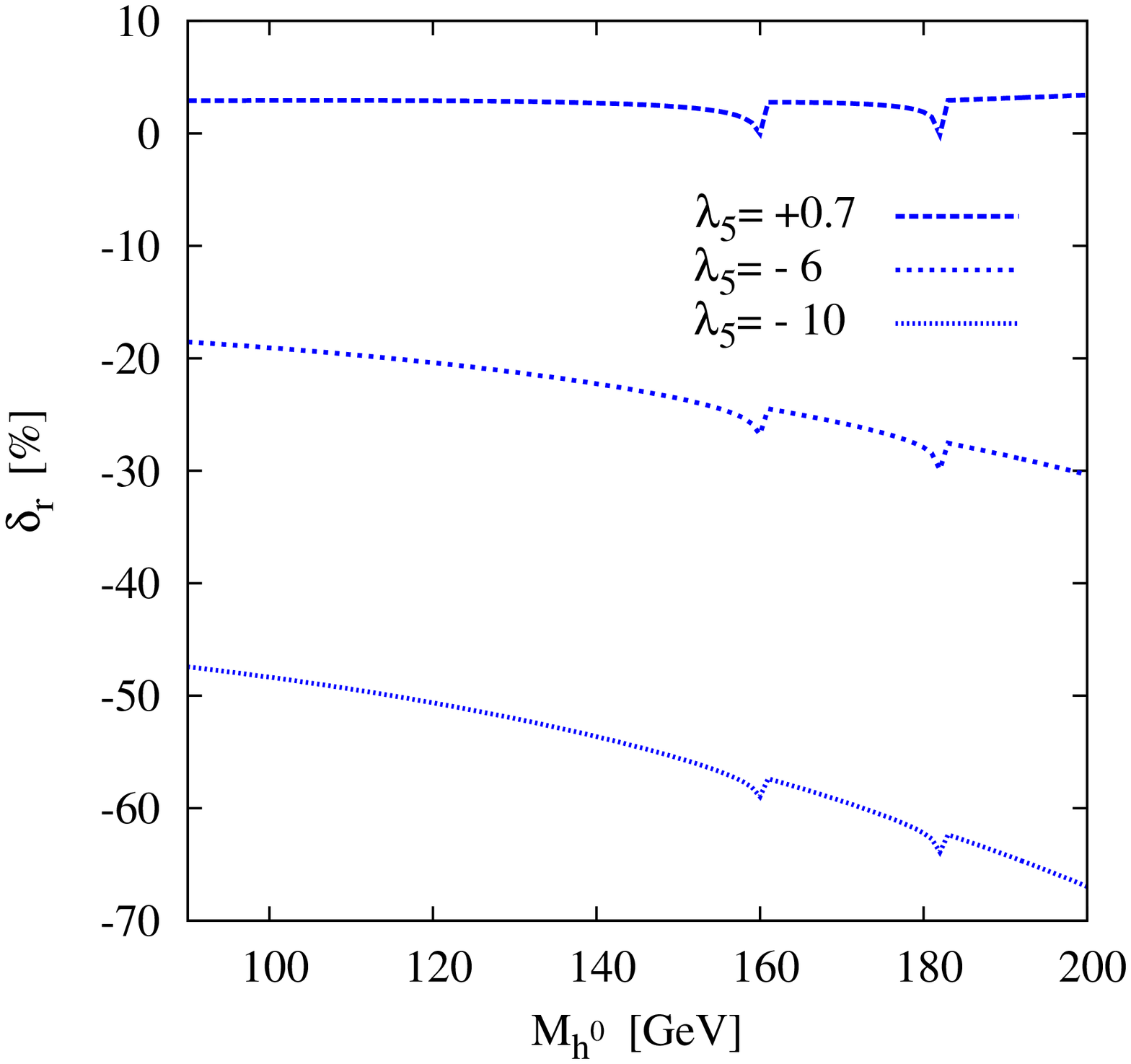} \\
&  &\\
\end{tabular}
\caption{Total cross section $\sigma(\APelectron\Pelectron \to \PHiggslightzero\PZ^0)$
(in fb, left panel) and relative one-loop correction $\delta_r$ (right panel)
at fixed $\sqrt{s}=500$ GeV as a function of $M_{\hzero}$
for Set B of Higgs boson masses,
cf. Table~\ref{tab:masses}. Shown are the results obtained within
three different values of $\lambda_5$
and for $\alpha=\beta-\pi/2$.}
\vspace{0.2cm}
\label{fig:plot4}
\end{center}
\end{figure*}

\begin{figure*}[ht!]
\begin{center}
\begin{tabular}{ccc}
\includegraphics[scale=0.37,angle=-90]{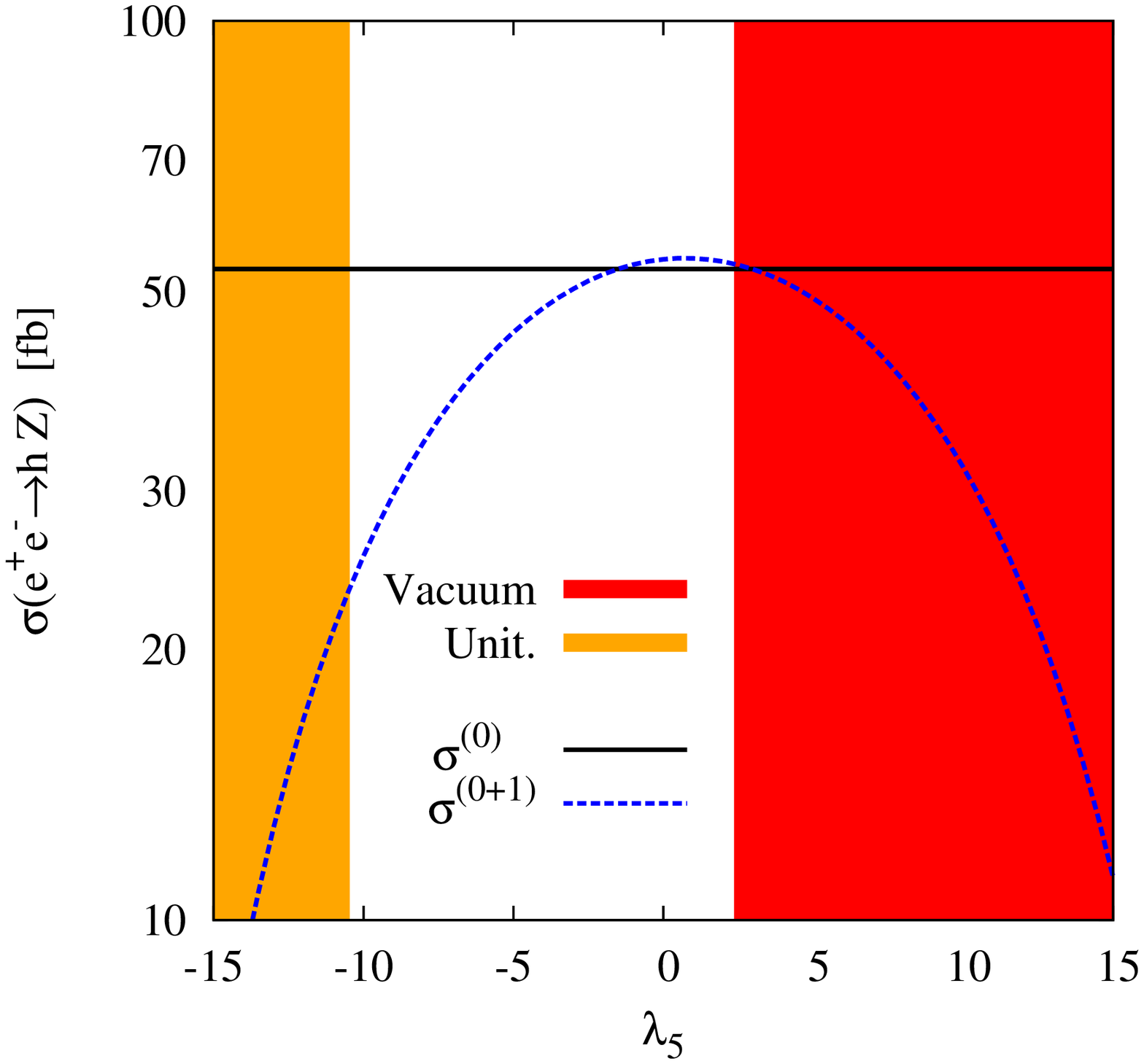} & \hspace{0.3cm} &
\includegraphics[scale=0.37,angle=-90]{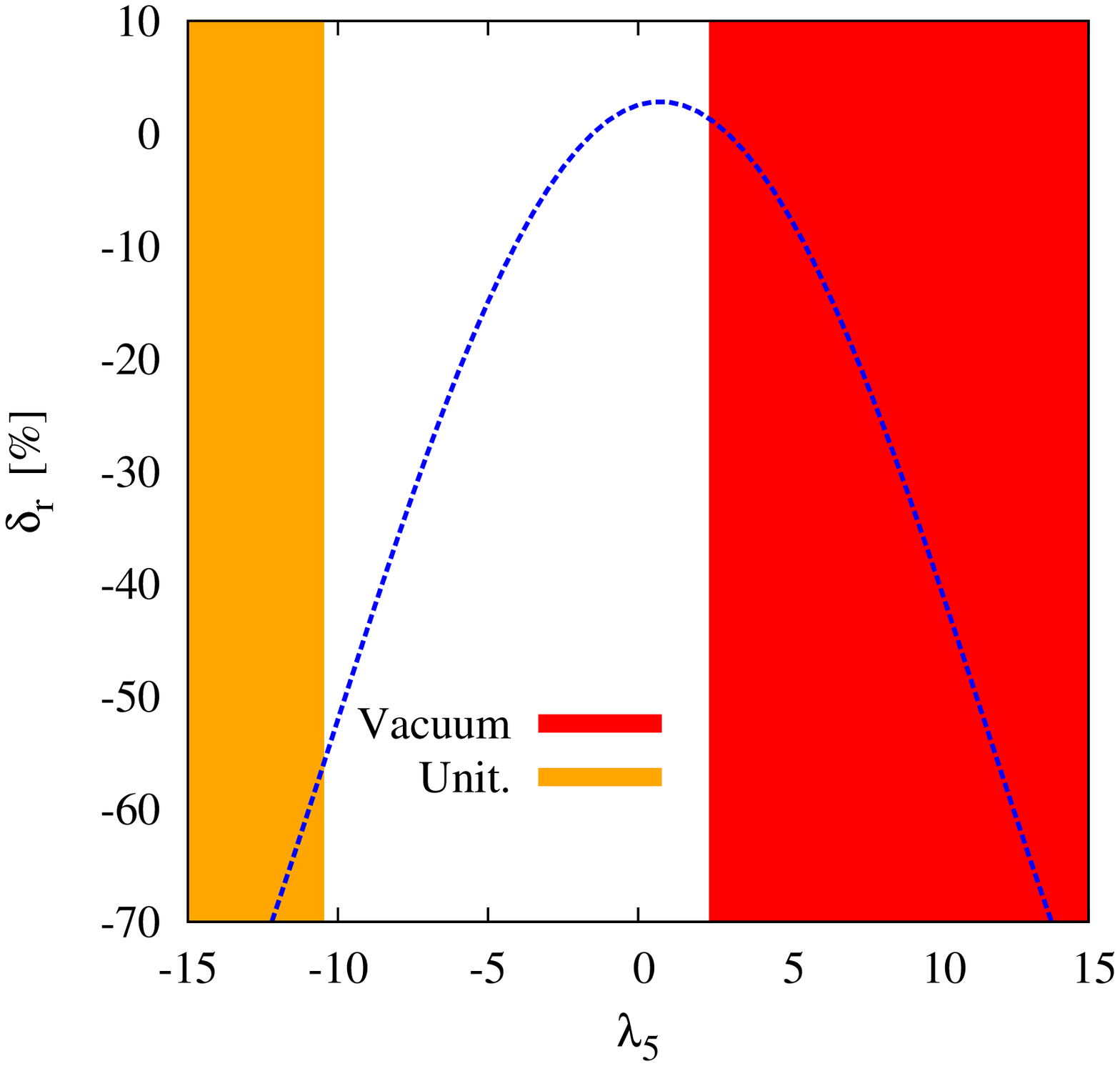}
\\
&  &\\
\end{tabular}
\caption{Total cross section $\sigma(\APelectron\Pelectron \to \PHiggslightzero\PZ^0)$
(in fb) at tree-level and at one-loop (left panel), together with
the relative one-loop correction $\delta_r$ (right panel, in $\%$) as a function of $\lambda_5$. The
results are obtained for $\tan\beta =1$, $\alpha=\beta-\pi/2$, $\sqrt{s}=500$ GeV and the
Higgs boson masses as in Set B of Table~\ref{tab:masses}.}
\vspace{0.2cm}
\label{fig:plot1}
\end{center}
\end{figure*}

In Figs.~\ref{fig:iso}-\ref{fig:plot1} we illustrate the fundamental
phenomenological features associated to the process $\phzero$ in the
scope of the general 2HDM. Fig.~\ref{fig:iso} summarizes the pattern
of radiative corrections $\delta_r$ projected onto the
$(\tan\beta,\,\lambda_5)$ and $(\sin\alpha\,,\tan\beta)$ planes, the
former at fixed $\alpha = \beta - \pi/2$ and the latter with
$\lambda_5 = -2$. We remark that for $\alpha = \beta-\pi/2$ the
$\hzero\PZ^0\PZ^0$ tree-level coupling takes on the SM form and
therefore is maximal. For definiteness, these plots have been
generated for a Higgs boson mass spectrum as in Set B (cf. Table~
\ref{tab:masses}), and at the fiducial ILC startup center-of-mass
energy,
 $\sqrt{s} = 500\,\GeV$.
Although the range $\tan\beta\gtrsim 1$ is usually the preferred
one from the theoretical point of view, we entertain the
possibility that $\tan\beta$ can be slightly below $1$ in order
to better assess the behavior around this value.
Fig.~\ref{fig:iso} illustrates that the allowed region in the
$(\tan\beta,\,\lambda_5)$ plane is severely restrained by the
theoretical constraints stemming from the perturbative unitarity
and vacuum stability. On the one hand, $\lambda_5>0$ values are
strongly disfavored by the vacuum stability condition; on the
other hand, the unitarity constraints tend to disfavor moderate
and large values of $\tan\beta$ (specially $\tan\beta$ values
significantly larger than $1$) as well as of $\tan\beta\ll 1$.
Altogether these constraints set an approximate lower bound of
$\lambda_5\sim -10$ for $\tan\beta \sim 1$ and a rigid upper
bound excluding almost all positive values of $\lambda_5$ for any
$\tan\beta$. The combined set of constraints builds up a
characteristic physical domain, with a valley-shaped area
centered at $\tan\beta \sim 1$ that sinks into the $\lambda_5 < 0$
region and becomes narrower with growing $|\lambda_5|$.

The fact that the curves of constant $\delta_r$ do not depend on
$\tan\beta$ (left panel of the Fig.\,\ref{fig:iso}) is related to
the choice $\alpha = \beta - \pi/2$, which we have made in order
to consider the situation where the tree-level cross-section for
$\hzero\PZ^0$ production is maximal. For this choice of the mixing
angle $\alpha$, it turns out that no particular enhancement shows
up for any value of $\tan\beta$, neither from the 3H
self-couplings nor from the Higgs-top quark Yukawa couplings.
Indeed, under the condition $\alpha = \beta - \pi/2$, the
$\hzero\APtop\Ptop$ Yukawa coupling does not depend on
$\tan\beta$:
\begin{eqnarray}
\lambda_{\hzero\APtop\Ptop}\,\Big|_{\alpha = \beta-\frac{\pi}{2}} =
 \frac{e\cos\alpha\,m_t}{2\,M_W\,\sw\sin\beta}
\,\Bigg|_{\alpha = \beta-\frac{\pi}{2}} =
-\,\frac{e\,m_t}{2\,M_W\,\sw}\,,
\nonumber \\
\label{eq:notanb1}
\end{eqnarray}
and at the same time the 3H self-couplings which are relevant for
this channel become independent of $\tan\beta$; notice, for example, that
\begin{eqnarray}
&&\lambda_{\hzero\Hzero\Hzero}\Big|_{\alpha = \beta - \pi/2} = \nonumber \\
&& \quad \frac{ie}{2\,M_W\,\sw}\,\left[(M^2_{\hzero} + 2\,M^2_{\Hzero}) -
\frac{4\lambda_5\,M_W^2\,\swd}{e^2} \right]\, . \nonumber \\
\label{eq:notanb2}
\end{eqnarray}
The final outcome is that the potential dependence of the
computed observables on $\tan\beta$ vanishes as long as we stick
to these $\alpha = \beta - \pi/2$ configurations -- in which the
tree-level coupling $\hzero\PZ^0\PZ^0$ is maximum and formally
equivalent to that of the SM \footnote{An equivalent discussion
would hold for the $\Hzero\PZ^0$ channel, under the complementary
condition $\alpha = \beta$.}. As a result, the only feasible
mechanism able to significantly enhance the quantum effects in
these scenarios is by increasing the value of the $|\lambda_5|$
parameter (towards more negative values, so as to be consistent
with vacuum stability).

Once we depart from the $\alpha = \beta-\pi/2$ setting, we recover
of course the expected dependence of the computed observables with
$\tan\beta$ (cf. right panel of Fig.~\ref{fig:iso}), but then the
lowest order $\hzero\PZ^0$ production cross section becomes smaller.
For $\tan\beta < 1$, radiative corrections are boosted as a result
of the enhanced 3H self-couplings, and partially also due to the
Higgs-top quark Yukawa couplings. Either way, their overall effect
is to suppress the tree-level signal, as such leading corrections
are triggered primordially by the finite Higgs boson WF corrections.
Another source of enhancement of $\delta_r$ appears near the region
where $\beta\sim\alpha$. However, this effect is not caused by a
real increment of the one-loop cross-section $\sigma^{(0+1)}$, but
by a mere suppression of the cross-section at the Born-level, i.e.
$\sigma^{(0)}$ in the denominator of Eq.\,(\ref{eq:deltar}), and in
this sense it is an uninteresting situation.

In Fig.~\ref{fig:plot3} we explore the evolution of the cross
section as a function of the center-of-mass energy. We include, in
each plot, the tree-level contribution $\sigma^{(0)}$ and the
corresponding loop-corrected value, $\sigma^{(0+1)}$, for different
values of $\lambda_5$. The right panel tracks the related behavior
of the quantum correction $\delta_r$ as a function of $\sqrt{s}$,
for the same set of $\lambda_5$ values. The plots are generated for
Set B of Higgs boson masses, at fixed $\tan\beta=1$ and
$\alpha=\beta-\pi/2$. The leading-order cross section $\sigma^{(0)}$
curve exhibits the expected behavior with $\sqrt{s}$, as it scales
with the $s$-channel $\PZ^0$-boson propagator, namely proportional
to $1/(s-M_Z^2)$. A similar pattern is also followed by the full
loop corrected cross-section.

The range where the relative one-loop correction $\delta_r$ is
positive is very reduced and it is confined to a regime where the
center-of-mass energy is around the startup value for the ILC
($\sqrt{s}\gtrsim 500$ GeV) and where $\lambda_5$ adopts the (small)
positive values allowed by the constraints. It should not come as a
surprise that this behavior is similar to the SM result found in
Fig. \ref{fig:plotSM}; indeed, being $\lambda_5$ small there can not
be significant 2HDM enhancements (not even from $\tan\beta$, which
is $1$) with respect to the corresponding SM process. Let us recall
that for Set B the maximum value of $\lambda_5$ allowed by the
vacuum stability bounds is $\lambda_5\sim 0.65$. Large negative
$\lambda_5$ values give rise to significant (negative) quantum
effects, and translate into loop-corrected cross sections
$\sigma^{(0+1)}$ depleted $40\%$ to $70\%$ with respect to the
tree-level predictions in the entire range from $\sqrt{s}=500$ GeV
to $\sqrt{s}=3$ TeV. Since the tree-level $\hzero\PZ^0\PZ^0$
coupling is equivalent to the SM coupling $\PHiggs\PZ^0\PZ^0$ for
$\alpha = \beta-\pi/2$, the right panel of Fig.~\ref{fig:plot3} also
illustrates the departure of the 2HDM loop-corrected cross section
with respect to the tree-level SM cross-section, namely
$(\sigma^{(0+1)}_{2HDM} - \sigma_{SM}^{(0)})/\sigma_{SM}^{(0)}$.

The behavior of $\sigma(\phzero)$ as a function of the Higgs boson
mass $M_{\hzero}$ is presented in Fig.~\ref{fig:plot4}. This figure
is the 2HDM counterpart of Fig.~\ref{fig:plotSM} corresponding to
the SM case. We superimpose the tree-level and the loop-corrected
cross-sections for the Set B of Higgs boson masses, by setting
$\tan\beta = 1$, $\alpha=\beta-\pi/2$ and using three different
values of $\lambda_5$. The center-of-mass energy is settled at the
fiducial value $\sqrt{s}=500$ GeV. Obviously, the raise of the Higgs
boson mass $M_{\hzero}$ implies a reduction of the available phase
space, so that the cross-section falls down. In the left panel of
Fig.~\ref{fig:plot4}, both the tree-level and the loop-corrected
cross-sections decrease monotonically with the growing of
$M_{\hzero}$. The $\PZ^0\PZ^0$ and $\PW^+\PW^-$ thresholds are also
barely visible therein. In the right panel of the same figure we
observe a steady increase of the negative value of the correction
for heavier Higgs boson mass, whereas the case with positive
correction remains almost stable.

Finally, in Fig.~\ref{fig:plot1} we display the tree-level
contribution and the total (one-loop corrected) cross section (left
panel), together with the relative radiative correction $\delta_r$
(right panel), as a function of the parameter $\lambda_5$. These
plots have been generated in the same benchmark conditions as in
Fig.~\ref{fig:plot4} and using a fiducial ILC start-up
center-of-mass energy $\sqrt{s}=500$ GeV. Let us remember that the
red shaded area on the right hand side of that figure (specifically
for $\lambda_5\gtrsim 2.4$) stands for the region excluded by the
vacuum stability bounds, whereas the light gray shaded area on the
left ($\lambda_5\lesssim -10.5$) signals the domain excluded by the
perturbative unitarity bounds. The cross section is seen to follow
the expected behavior, which we have derived from the estimate of
the leading effect -- cf. Eq.\, \eqref{eq:demo3} -- in combination
with the structure of the trilinear coupling (\ref{eq:notanb2}),
{namely a negative quadratic dependence of the scattering amplitude,
which transates into $\sigma(\APelectron\Pelectron \to
\PHiggslightzero\PZ^0)\sim (a-b\lambda^2_5)^2$, emerging ultimately
from the finite WF corrections to the external Higgs boson leg
(notice the inclusion of the leading quartic corrections)}. For
moderate negative $\lambda_5$, the relative size of the one-loop
quantum corrections can reach up to $\sim -50\%$. Positive
$\delta_r$, however, can only take place for $\lambda_5$ close to
zero.
\begin{table*}[tbh]
\begin{center}
\begin{tabular}{|c|r||c|c|c||c|c|c|} \hline
\multicolumn{2}{|c||}{} & \multicolumn{3}{|c||}{$\hzero\PZ^0$} & \multicolumn{3}{|c|}{$\Hzero\PZ^0$}\\\cline{3-8}
\multicolumn{2}{|c||}{} & $\alpha=\beta-\pi/2$ & $\alpha=\beta-\pi/3$ & $\alpha=\beta-\pi/6$ & $\alpha=\beta-\pi/3$ & $\alpha=\beta-\pi/6$ & $~~~\alpha=\beta~~~$\\\hline\hline
\multirow{2}{1.2cm}{Set A}& $\sigma_{max}\,[\femtobarn]$ & $ 42.61$ & $ 27.89$ & $  8.17$ & $  5.36$ & $ 17.86$ & $ 10.70$ \\\cline{2-8}
                          & $\delta_r\,[\%]$             & $-22.79$ & $-32.62$ & $-40.76$ & $-40.92$ & $-34.40$ & $-70.51$ \\\hline\hline
\multirow{2}{1.2cm}{Set B}& $\sigma_{max}\,[\femtobarn]$ & $ 24.80$ & $ 18.84$ & $  7.24$ & $  7.05$ & $ 15.51$ & $ 14.99$ \\\cline{2-8}
                          & $\delta_r\,[\%]$             & $-53.18$ & $-52.57$ & $-45.29$ & $-41.19$ & $-56.87$ & $-68.74$ \\\hline\hline
\multirow{2}{1.2cm}{Set C}& $\sigma_{max}\,[\femtobarn]$ & $ 28.81$ & $ 21.97$ & $  7.75$ & $  7.83$ & $ 20.82$ & $ 25.60$ \\\cline{2-8}
                          & $\delta_r\,[\%]$             & $-43.91$ & $-42.97$ & $-39.66$ & $-36.96$ & $-44.13$ & $-48.48$ \\\hline\hline
\multirow{2}{1.2cm}{Set D}& $\sigma_{max}\,[\femtobarn]$ & $ 40.77$ & $ 29.25$ & $  9.30$ & $  8.56$ & $ 23.95$ & $ 28.59$ \\\cline{2-8}
                          & $\delta_r\,[\%]$             & $-17.95$ & $-21.52$ & $-25.15$ & $-15.14$ & $-20.84$ & $-29.12$ \\\hline

\end{tabular}
\caption{Maximum total cross section
$\sigma^{(0+1)}(\APelectron\Pelectron \to h\PZ^0)$, for $h =
\hzero,\Hzero$, at $\sqrt{s} = 500\,\GeV$, together with the
relative size of the radiative corrections $\delta_r$, for the
different sets of Higgs bosons masses quoted in
Table~\ref{tab:masses}. The results are obtained at fixed $\tan\beta
= 1$ and different values of $\alpha$, with $\lambda_5$ at its
largest negative attainable value, namely: $\lambda_5 \simeq -9$
(for Set A), $\lambda_5 \simeq -10$ (Sets B,C) and $\tan\beta\simeq
-8$ (Set D). } \label{tab:xs500}
\end{center}
\end{table*}

A rather comprehensive survey of the predicted cross-sections for
different values of the tree-level coupling and different Higgs mass
setups is presented in Table~\ref{tab:xs500}. We display the results
for the one-loop corrected cross-sections (in fb), together with the
relative radiative corrections (cf. Eq.~\eqref{eq:deltar}), for both
neutral Higgs-strahlung channels $\hzero\PZ^0$ and $\Hzero\PZ^0$.
Once more, we set $\sqrt{s} = 500\,\GeV$; and we work at $\tan\beta
= 1$ and maximum allowed $|\lambda_5|$, since we are mostly
interested in spotlighting the imprints of the 3H self-couplings in
the quantum effects associated to the Higgs-strahlung mechanism. Let
us also recall in passing that $\alpha = \beta -\pi/2$ maximizes the
tree-level $\hzero\PZ^0\PZ^0$ coupling, whilst $\Hzero\PZ^0\PZ^0$ is
optimal in the complementary regime, $\alpha = \beta$. From the
table we may read out that radiative corrections in these regimes
are certainly large, regardless of the details of the chosen mass
spectrum, the tree-level coupling and the actual channel under
consideration. In a nutshell: the characteristic signature of the
enhanced 3H self-couplings in the pattern of quantum effects on the
Higgs-strahlung processes is rather universal and manifests in the
form of a substantial depletion of the tree-level signal (typically
in the range of $\delta_r \sim -20 / -60\,\%$). Fortunately, even
under such a dramatic suppression the final loop-corrected cross
sections stay at the level of a few tens of fb -- thereby amounting
to a non-negligible yield of $\sim 10^3 - 10^4$ events per
$500\,\invfb$.

A note of caution should be given at this point in regard to some
particular configurations of the $\Hzero\PZ^0$ channel. It turns out
that certain choices of Higgs boson masses (e.g. for MSSM-like mass
splittings) may lie close, or simply beyond, the kinematical
threshold for the decay process $\Hzero \to \hzero\hzero$; that is
to say, $M_{\Hzero} \simeq 2\,M_{\hzero}$. If we then consider a
regime wherein $\alpha = \beta$ and sizable $|\lambda_5|$, then the
\Hzero\,  WF-correction undergoes a remarkable boost, which results
from the combination of two independent sources of enhancement,
namely i) the actual strength of the $\Hzero\hzero\hzero$ coupling
-- which is maximally enlarged in this regime; and ii) the
kinematical enhancement due to the vicinity of the $\Hzero \to
\hzero\hzero$ threshold, which is reflected as a sharp peak in
$\delta\,Z_{\Hzero} \sim
B_0'(M^2_{\Hzero},M^2_{\hzero},M^2_{\hzero})$. However, these
$\Hzero$ WF-corrections become so large that the perturbative
formula~\eqref{eq:demo3} is no longer valid and we have to keep the
original (unexpanded) expression $\hat{Z}^{1/2}_{\Hzero} = \left[ 1
+ {\Re e\retildehat_{\Hzero}'(M^2_{\Hzero})} \right]^{-1/2}$.
\noindent Out of such corner in the parameter space, the
higher-order effects involved in the above resummed form of the
WF-corrections become harmless and totally inconspicuous and can be
safely discarded. If, alternatively, we were considering a mass
setup such that $M_{\Hzero} > 2 M_{\hzero}$, then the decay $\Hzero
\to \hzero\hzero$ would be open and, in the scenario of maximum
$\Hzero\hzero\hzero$ coupling, it would furnish a very large width
for $\Hzero$. In this particular setup, the process would
effectively boil down to the double Higgs-strahlung channels
$\APelectron\Pelectron \to \Hzero\PZ^0 \to \hzero\hzero\PZ^0$
previously explored in the literature~\cite{arhrib}.

The following comment is in order to clarify the role played by the
remaining (potential) sources of enhancement. We have seen that the
various constraints restrict very significantly the range of allowed
values of $\tan\beta$, and enforce it to stay around $1$ when
$|\lambda_5|$ is maximum in the region $\lambda_5<0$. As a result,
the contributions from the top quark and bottom quark Yukawa
couplings cannot be augmented in the domain where the trilinear
couplings are maximal. Remarkably enough, let us also emphasize that
they cannot be enhanced even in the region where $|\lambda_5|$ is
small or zero. To see why, notice that although in such region the
parameter $\tan\beta$ can be very large or small and still be
compatible with the constraints (cf. Fig.\,\ref{fig:iso}), then all
the terms in the trilinear couplings (cf. Table II of
Ref.\,\cite{main}) which are not proportional to $\lambda_5$ become
of order one (i.e. cannot be promoted to high values) in the regime
where the tree-level production cross-section for $\hzero$ or
$\Hzero$ is optimal. Not only so, in the very same regime the Yukawa
couplings of both the top and bottom quarks cannot be enhanced
either. One can easily check all these features explicitly by
observing that, in the regions where the corresponding tree-level
processes are maximal, the $\tan\beta$-enhancements are canceled in
all the couplings. We have checked numerically that the residual
corrections (positive and negative) are merely of a few percent for
any value of $\tan\beta$. At the end of the day, we conclude that
the only sizeable and eventually measurable quantum effects on the
processes under study are those stemming potentially from the
enhancement of the $\lambda_5$ coupling, not from the Yukawa
couplings.

Let us finally emphasize that the complementarity between the
neutral final states $\hzero\PZ^0$/$\hzero\Azero$ and
$\Hzero\PZ^0$/$\Hzero\Azero$, i.e. processes (\ref{2H}) and
(\ref{eq:hz}), is a unique chance for analyzing potentially big
correlations between quantum effects in the 2HDM production cross
sections, as there is no similar opportunity in the charged sector.
Indeed, the final state charged counterparts $\Hpm \PW^{\mp}$ are
suppressed owing to the fact that the $Z \Hpm \PW^{\mp}$ vertices
are forbidden at the tree-level in any 2HDM extension of the
SM\,\cite{Grifols:1980uq} and therefore they can only be studied in
more general extensions of the Higgs sector\,\cite{Grifols:1980ih},
or through loop-induced $\Hpm \PW^{\mp}$ vertices in the
2HDM\,\cite{Arhrib:1999rg} and in the MSSM\,\cite{Brein:2006xda}. In
these loop-induced mechanisms, the cross section for the associated
$\Hpm \PW^{\mp}$ production in a linear collider is rather meager --
generally below $1$ fb at the startup ILC energy (for charged Higgs
masses comparable to the ones we have considered for the
$\hzero\PZ^0$/$\hzero\Azero$ production) and within the region
allowed by the current constraints. We therefore deem quite
difficult to use the $\Hpm \PW^{\mp}$ channel to extract additional
information. In our opinion, the main task should be focused on
performing precision tests of the $\hzero\PZ^0$/$\hzero\Azero$ final
states, together with the $\Hzero\PZ^0$/$\Hzero\Azero$ ones (if the
mass of the heavy $\CP$-even Higgs is not too large).

\section{Discussion and conclusions}
\label{sec:conclusions}

In this article, we have concentrated on the analysis of the
production of neutral Higgs bosons in association with the $\PZ^0$
gauge boson at the future linac facilities within the framework of
the general (non-supersymmetric) Two-Higgs-Doublet-Model (2HDM). Our
basic endeavor has been twofold: i) on the one hand to study the
impact of radiative corrections to the final predicted rates; and
ii) on the other hand to correlate such quantum effects with the
enhancement potential of the 3H self-interactions, which are a
genuine dynamical feature of the 2HDM -- in the sense that it is
unmatched to its supersymmetric counterpart. The upshot of our
analysis singles out sizable (although negative) quantum effects
which are correlated to the enhancement properties of the 3H
self-interactions in the general 2HDM. Such large effects are
identified fundamentally in the region of parameter space with
$\tan\beta \simeq 1$ and where the coupling $|\lambda_5|$ is at its
maximum possible value compatible with the various constraints. The
quantum effects reach typically $\delta\sigma/\sigma \sim -20\% /
-60 \%$, for $|\lambda_5|\sim 8/\sim 10$ ($\lambda_5 < 0$), this
being the crucial parameter that tunes the actual size of the 3H
self-couplings. Let us stress that the most stringent limits on
$\lambda_5$ are placed by the conditions of perturbative unitarity
and vacuum stability\footnote{Recently, a combined analysis of
different $B$-meson physics constraints over the 2HDM parameter
space suggests that values of $\tan\beta \sim 1$ could be disfavored
for charged Higgs masses $M_{\Hpm}$ too near to the lowest mass
limit of $300$ GeV (and certainly below) \,\cite{mahmoudi}. However,
the level of significance is not high and, in addition, our leading
quantum corrections are basically insensitive to the charged Higgs
mass. Therefore, a shift of $M_{\Hpm}$ slightly upwards restores the
possibility of $\tan\beta \sim 1$ at, say, $2\sigma$ without
altering significantly our results (as we have explicitly
checked).}.

Although the vertex corrections to the $\hzero\PZ^0\PZ^0 /
\Hzero\PZ^0\PZ^0$ couplings are sensitive to such 3H
self-interactions through Higgs-mediated one-loop diagrams, the most
significant quantum effects are induced by the wave-function
renormalization corrections to the $\hzero/\Hzero$ Higgs boson
external lines. These (negative) effects triggered by the trilinear
couplings are of order $\alpha_{\rm ew}\lambda_{3H}^2$, and are
therefore very responsive to changes in the value of the parameter
$\lambda_5$ in the Higgs potential of the 2HDM. In turn, the
gauge-boson and fermion one-loop contributions remain subleading
(including the effects from the Yukawa couplings) as they show no
remarkable departures from their SM analogues in the relevant
regions of parameter space -- which yield $\delta \sigma/\sigma$ at
the level of a few percent. Most important is also the fact that the
combined analysis of such Higgs-strahlung events with the previously
considered Higgs-pair production processes ($\hzero\Azero$,
$\Hzero\Azero$) -- see \cite{main} for details -- could be
instrumental as a highly sensitive probe of the underlying
architecture of the Higgs sector. At the fiducial center-of-mass
energy value of $\sqrt{s} = 500\,\GeV$, and when the genuine
enhancement mechanisms of the 2HDM are active, the $h\Azero$ events
are remarkably strengthened at the one-loop level while the $h\PZ^0$
channels are simultaneously depleted.

We can assert that the described pattern of leading quantum effects
emerges as a kind of  ``universal'' feature of the 2HDM as far as
the predictions for the Higgs-strahlung processes are concerned,
meaning that this pattern is virtually independent of the details of
the Higgs mass spectrum and of the actual production channel
($\hzero\PZ^0$, $\Hzero\PZ^0$). The same is true for the pairwise
production channels\,\cite{main}.  Therefore they all can be
physically significant provided the produced Higgs boson is not too
heavy for the actual center-of-mass energy. Focusing on the
Higgs-strahlung processes, the rise of large (and negative)
radiative corrections to their cross-sections within the 2HDM can be
regarded as a characteristic imprint of a general two-Higgs-doublet
model structure. Furthermore, the presence of a $\PZ^0$-boson in the
final state (and hence of its clear-cut leptonic signature $\PZ^0
\to l^+ l^-$) should enable a rather comfortable tagging of these
Higgs events following a method entirely similar to the original
Bjorken mechanism\,\cite{Bjorken76} in the SM. Although the
corresponding cross-sections at the higher operating energies
planned for the future linear colliders are smaller (as compared to
LEP), they are nevertheless sufficiently sizeable (typically in the
range $10-40$ fb) for a relatively comfortable practical measurement
at the higher luminosities scheduled for these machines.  For
example, the associated decay of the accompanying neutral \CP-even
Higgs-boson would manifest basically in the form of either i)
collimated and highly-energetic $b$-quark or $\tau$-lepton jets, for
$M_{\hzero}\lesssim 2\,M_V \sim 180$ GeV; and ii) $\hzero,\Hzero \to
\PW^+\PW^- \to l^+ l^-$ + missing energy; or $\Hzero \to \PZ^0 \PZ^0
\to l^+ l^- l'^+ l'^- $, for $M_{\Hzero}
> 2\,M_V$ (in the optimal regime for the tree-level process).

Some discussion on how the corresponding MSSM Higgs boson production
cross-sections compare to the 2HDM ones seems also appropriate. In
the MSSM, the maximum quantum effects on $h\PZ^0$ production are
typically milder. Here, the dominance of the WF-corrections is also
the main source of one-loop effects\,\cite{mssmloop3}. They are
usually reabsorbed into the tree-level $\PZ\PZ h$ couplings
(\ref{eq:lag0}), specifically in the $\CP$-even mixing angle
$\alpha$, which then becomes an effective mixing parameter.
Consider, for example, the behavior of the characteristic
trigonometric coupling $\cos^2(\beta-\alpha)$ in the $\PZ\PZ\Hzero$
interaction vertex, which in the MSSM dies away with growing values
of $M_{\Azero}$ (both at the tree-level and at one-loop). It follows
that in a situation where $M_{\Azero}$ is sufficiently heavy, say
above $200$ GeV, the MSSM would predict measurable rates for the
channel $\hzero\PZ^0$ (and also for $\Hzero\Azero$, if $\sqrt{s}$ is
sufficiently high), whereas the complementary channels ($\Hzero\PZ^0
/ \hzero\Azero$) would be virtually below the observability
threshold. In contrast, in this particular SM-like regime (in which
$\sin(\beta-\alpha) \to 1$) the $\hzero\PZ^0$ cross-section in the
2HDM could well be exhibiting the trademark suppression induced at
one-loop by large 3H self-couplings.

Furthermore, in the MSSM, the dynamical origin of the leading
WF-effects does not reside in the Higgs boson self-couplings, but on
the Yukawa couplings with fermions and also in the large Yukawa-like
couplings of Higgs bosons with sfermions, particularly with the
stop. Thus, in both models (2HDM or MSSM) the main source of quantum
effects emanates from the renormalization of the Higgs boson
external lines, but the kind of interactions involved in each case
is radically different. On the quantitative side, the impact of the
quantum effects in the MSSM case turns out to be considerably milder
and as a consequence the cross-sections for Higgs-strahlung
production remain similar to the SM cross-section (i.e. for the
Bjorken process). The corresponding 2HDM cross-sections, instead,
can be significantly smaller -- if taking the same Higgs boson
masses (e.g. Sets A and B of Table ~\ref{tab:masses}, which mimic
the MSSM Higgs boson mass spectrum) owing to the aforementioned
large suppression effect from the Higgs boson WF renormalization. At
the same time, as we have already mentioned, the cross-sections for
neutral Higgs boson pair production in the general 2HDM -- cf.
Eq.\,(\ref{2H}) -- become substantially larger than the MSSM
counterparts (for the same Higgs boson mass spectrum), carrying very
sizeable and positive quantum effects\,\cite{main}. Therefore, an
interesting combined picture emerges in the 2HDM context, in which a
large ($\sim 50\%$) enhancement of the pairwise neutral Higgs boson
production is simultaneously accompanied by a drastic drop (of
similar size) of the Higgs-strahlung events. As this situation is
completely impossible to realize in the MSSM, this feature could be
used as a strong characteristic signature to discriminate between
these processes in the general 2HDM and in the MSSM. In other words,
it could provide an essential quantum handle enabling us to perform
a proper identification of the kind of Higgs bosons produced in a
linear collider.

In summary, we have analyzed the classical Higgs-boson strahlung
processes at linear colliders in the light of the general 2HDM, and
we have elucidated very significant quantum imprints which
spotlight, once more, the stupendous phenomenological possibilities
that triple Higgs boson self-interactions could encapsulate in
non-supersymmetric extended Higgs sectors. To be sure, regardless of
whether the LHC is finally capable to discover the Higgs boson(s), a
paramount effort will still be mandatory in order to completely
settle its experimental basis and to uncover the nature of the
spinless constituents behind the electroweak symmetry-breaking
mechanism. As we have shown, experiments at the future linear
collider facilities could play a crucial role in this momentous
endeavor.

\vspace{0.3cm}
 \noindent
\textbf{Acknowledgments}\, NB thanks an {ER} position of the EU
project MRTN-CT-2006-035505 {HEPTools} and the hospitality at the
Dept. ECM of the Univ. de Barcelona; DLV acknowledges the MEC FPU
grant AP2006-00357. The work of JS has been supported in part by MEC
and FEDER under project FPA2007-66665 and by DIUE/CUR Generalitat de
Catalunya under project 2009SGR502 and by the Spanish
Consolider-Ingenio 2010 program CPAN CSD2007-00042. DLV wishes to
thank the hospitality of the Theory Group at the Physikalisches
Institut of the University of Bonn and the Bethe Center for
Theoretical Physics. Discussions with Karina E. Williams are
gratefully acknowledged.

\newcommand{\JHEP}[3]{ {JHEP} {#1} (#2)  {#3}}
\newcommand{\NPB}[3]{{\sl Nucl. Phys. } {\bf B#1} (#2)  {#3}}
\newcommand{\NPPS}[3]{{\sl Nucl. Phys. Proc. Supp. } {\bf #1} (#2)  {#3}}
\newcommand{\PRD}[3]{{\sl Phys. Rev. } {\bf D#1} (#2)   {#3}}
\newcommand{\PLB}[3]{{\sl Phys. Lett. } {\bf B#1} (#2)  {#3}}
\newcommand{\PL}[3]{{\sl Phys. Lett. } {#1} (#2)  {#3}}
\newcommand{\EPJ}[3]{{\sl Eur. Phys. J } {\bf C#1} (#2)  {#3}}
\newcommand{\PR}[3]{{\sl Phys. Rep } {\bf #1} (#2)  {#3}}
\newcommand{\RMP}[3]{{\sl Rev. Mod. Phys. } {\bf #1} (#2)  {#3}}
\newcommand{\IJMP}[3]{{\sl Int. J. of Mod. Phys. } {\bf #1} (#2)  {#3}}
\newcommand{\PRL}[3]{{\sl Phys. Rev. Lett. } {\bf #1} (#2) {#3}}
\newcommand{\ZFP}[3]{{\sl Zeitsch. f. Physik } {\bf C#1} (#2)  {#3}}
\newcommand{\MPLA}[3]{{\sl Mod. Phys. Lett. } {\bf A#1} (#2) {#3}}
\newcommand{\JPG}[3]{{\sl J. Phys.} {\bf G#1} (#2)  {#3}}
\newcommand{\FP}[3]{{\sl Fortsch. Phys.} {\bf G#1} (#2)  {#3}}
\newcommand{\PTP}[3]{{\sl Prog. Theor. Phys. Suppl.} {\bf G#1} (#2)  {#3}}

\end{document}